%
%A SET OF MACROS FOR WRITING PAPERS.
%
%\today GIVES TODAY'S DATE.
%
\def\today{\ifcase\month\or January\or February\or March\or April\or May\or
June\or July\or August\or September\or October\or November\or December\fi
\space\number\day, \number\year}
%
%\note{footnote} GIVES SEQUENTIALLY NUMBERED FOOTNOTES.
%
\newcount\notenumber

\def\note{\global\advance\notenumber by 1 \footnote{$^{\the\notenumber}$}}
%
%\numbereq SEQUENTIALLY NUMBERS EQUATIONS ON THE RIGHT (number)
%
\newif\ifsectionnumbering
\newcount\eqnumber
\def\cleareqnumber{\eqnumber=0}
\def\numbereq{\global\advance\eqnumber by 1
\ifsectionnumbering\eqno(\the\secnumber.\the\eqnumber)\else\eqno
(\the\eqnumber) \fi}
\def\eqalinno{{\global\advance\eqnumber by 1}
\ifsectionnumbering(\the\secnumber.\the\eqnumber)\else(\the\eqnumber)\fi}
\def\name#1{\ifsectionnumbering\xdef#1{\the\secnumber.\the\eqnumber}
\else\xdef#1{\the\eqnumber}\fi}

\sectionnumberingtrue
%
%\ref{\name} GIVES SEQUENTIALLY NUMBERED
%REFERENCES [number], AND ASSIGNS
%THAT NUMBER TO A MACRO \name AND WRITES REF. TO FILE 1.
%
\newcount\refnumber

\immediate\openout1=refs.tex
\immediate\write1{\noexpand\frenchspacing}
\immediate\write1{\parskip=0pt}
\def\ref#1#2{\global\advance\refnumber by 1%
[\the\refnumber]\xdef#1{\the\refnumber}%
\immediate\write1{\noexpand\item{[#1]}#2}}
\def\tie{\noexpand~}

%
% NEW SECTION: \newsection The Method. (terminate with a .)
%
\font\twelvebf=cmbx10 scaled \magstep1
\newcount\secnumber

\def\newsection#1.{\ifsectionnumbering\cleareqnumber\else\fi%
	\global\advance\secnumber by 1%
	\bigbreak\bigskip\par%
	\line{\twelvebf \the\secnumber. #1.\hfil}\nobreak\medskip\par\noindent}
%
%
%\Box GIVES WAVE OPERATOR, OR LAPLACIAN
%
\def \sqr#1#2{{\vcenter{\vbox{\hrule height.#2pt
	\hbox{\vrule width.#2pt height#1pt \kern#1pt
		\vrule width.#2pt}
		\hrule height.#2pt}}}}

%
%
%\twocolumns GIVES TWO-COLUMN OUTPUT
%
\newdimen\fullhsize
\def\fiddle{\fullhsize=6.5truein \hsize=3.2truein}
\def\fullline{\hbox to\fullhsize}
\def\mkhdline{\vbox to 0pt{\vskip-22.5pt
	\fullline{\vbox to8.5pt{}\the\headline}\vss}\nointerlineskip}
\def\mkftline{\baselineskip=24pt\fullline{\the\footline}}
\let\lr=L \newbox\leftcolumn
\def\twocolumns{\fiddle
	\output={\if L\lr \global\setbox\leftcolumn=\columnbox
		\global\let\lr=R \else \doubleformat \global\let\lr=L\fi
		\ifnum\outputpenalty>-20000 \else\dosupereject\fi}}
\def\doubleformat{\shipout\vbox{\mkhdline
		\fullline{\box\leftcolumn\hfil\columnbox}
		\mkftline} \advancepageno}
\def\columnbox{\leftline{\pagebody}}
%\nosectionnumbering
\magnification=1200
\def\pr#1 {Phys. Rev. {\bf D#1\tie }}
\def\pe#1 {Phys. Rev. {\bf #1\tie}}
\def\pre#1 {Phys. Rep. {\bf #1\tie}}
\def\pl#1 {Phys. Lett. {\bf #1B\tie }}
\def\prl#1 {Phys. Rev. Lett. {\bf #1\tie }}
\def\np#1 {Nucl. Phys. {\bf B#1\tie }}
\def\ap#1 {Ann. Phys. (NY) {\bf #1\tie }}
\def\cmp#1 {Commun. Math. Phys. {\bf #1\tie }}
\def\imp#1 {Int. Jour. Mod. Phys. {\bf A#1\tie }}
\def\mpl#1 {Mod. Phys. Lett. {\bf A#1\tie}}
\def\jhep#1 {JHEP {\bf #1\tie}}
\def\nuo#1 {Nuovo Cimento {\bf B#1\tie}}
\def\ussr#1 {Zh. Eksp. Teor. Fiz {\bf #1\tie }}
\def\jetp#1 {Soviet Physics JETP {\bf #1\tie }}
\def\rmp#1 {Rev. Mod. Phys. {\bf #1\tie }}
\def\tie{\noexpand~}
\def\ov{\bar}

\parskip=15pt plus 4pt minus 3pt
\headline{\ifnum \pageno>1\it\hfil Ginzburg-Landau Free Energy
Functional $\ldots$\else \hfil\fi}
\font\title=cmbx10 scaled\magstep1
\font\tit=cmti10 scaled\magstep1
\footline{\ifnum \pageno>1 \hfil \folio \hfil \else
\hfil\fi}
\raggedbottom

%%%%%%%%%%%%%%%%%%%%%%%
% jtl definitions

\overfullrule0pt

%%%%%%%%%%%%%%%%%%%%%%%

\rightline{\vbox{\hbox{RU01-12-B}\hbox{hep-th/0108256}}}
\vfill
\centerline{\title THE GINZBURG-LANDAU FREE ENERGY FUNCTIONAL}
\centerline{\title  OF COLOR SUPERCONDUCTIVITY AT WEAK COUPLING}
\vfill
{\centerline{\title Ioannis Giannakis${}^{a}$
and Hai-cang Ren${}^{a}$ \footnote{$^{\dag}$}
{\rm e-mail: \vtop{\baselineskip12pt
\hbox{giannak@theory.rockefeller.edu, ren@theory.rockefeller.edu,}}}}
}
\medskip
\centerline{$^{(a)}${\tit Physics Department, The Rockefeller
University}}
\centerline{\tit 1230 York Avenue, New York, NY
10021-6399}
\vfill
\centerline{\title Abstract}
\bigskip
{\narrower\narrower
We derive the Ginzburg-Landau free energy functional
of color superconductivity in terms of the thermal
diagrams of QCD in its perturbative region.
The zero mode of the quadratic term coefficient yields the same
transition temperature, including the pre-exponential factor,
as the one obtained previously from the Fredholm determinant of the two
quark scattering amplitude. All coefficients of the free energy
can be made identical to those of a BCS model by setting
the Fermi velocity of the latter equal to the speed of light.
We also calculate the induced symmetric color condensate
near $T_c$ and find that it scales as the cubic power
of the dominant antisymmetric color component. We show that
in the presence of
an inhomogeneity and a nonzero gauge potential,
while the color-flavor locked condensate dominates
in the bulk, the unlocked condensate, the octet, emerges
as a result of a simultaneous
color-flavor rotation in the core region of a
vortex filament or at the junction of super and normal phases.
\par}
\vfill\vfill\break

%%%%%%%%%%%%%%%%%%%%%%%

\newsection Introduction.%

The properties of hadronic matter under unusual
conditions, high temperature or high baryon density,
have been the subject of intense theoretical investigation,
especially during the last few years. Relativistic heavy-ion collision
experiments probe the
quark-gluon plasma both at nonzero temperature and nonzero 
chemical potential.
Quark matter at high 
baryon density and low temperature is expected to exist
inside neutron stars. The exploration of these exotic states 
of matter will hopefully
shed light on many non-perturbative aspects of
quantum chromodynamics.

Although great progress has been
achieved in our understanding of QCD
along the temperature axis in the two-dimensional
phase diagram of QCD with respect to temperature and
chemical potential,
based primarily in our ability to perform simulations
on the lattice, the problem of low temperature, high
density hadronic matter is less well understood mainly
because the fermion sign problem
at nonzero chemical potential makes numerical simulations
practically impossible at high baryon density. Thus, many issues of
the physics near the chemical potential axis can only be addressed
analytically.

At ultra high chemical potential, $\mu>>\Lambda_{\rm{QCD}}$, the 
interaction between two quarks near their Fermi surface is dominated 
by one gluon exchange because of the asymptotic freedom. Such an 
interaction can be decomposed into a color symmetric channel and a 
color antisymmetric channel. For a pair of quarks within the color
antisymmetric channel propagating in opposite directions, both
color-electric and color-magnetic parts of the interaction are attractive.
Therefore the Fermi surface is unstable against the formation of 
Cooper pairs and the quark matter becomes superconducting below a 
certain temperature \ref{\barrois}{B. Barrois, \np129(1977) 390, 
S. Frautschi, {\it Proceedings of the
Workshop on Hadronic Matter at Extreme Energy Density},
N. Cabibbo, Editor, Erice, Italy, 1978, D. Bailin
and A. Love, \pre107 (1984) 325, and references
therein for early works.}. Such a 
color superconducting state is expected to survive at medium high
chemical potential, $\mu>\Lambda_{\rm{QCD}}$ ($\mu=400\sim 500$MeV),
which corresponds to the baryon density achieved in RHIC and in
the core of a typical neutron star. The instanton interaction may also
contribute to the condensation at medium high chemical potential
\ref{\sca}{T. Schafer and E. V. Shuryak, \rmp70 (1998) 323.}.

Theoretically, there are two approaches which explore the physics of
the color superconductivity, one that
utilizes an effective action
involving only quark degrees of freedom while the other uses the
fundamental QCD action. Using the first approach, important progress
has been made by Alford, Rajagopal and Wilczek, who recognized that the
color-flavor locked condensate represents the
minimum of the bulk free energy and thus characterizes the super phase
\ref{\alford}{M. Alford, K. Rajagopal and F.
Wilczek, \np537 (1999) 443, M. Alford, hep-th/0102047,
K. Rajagopal and Wilczek, hep-ph/0011333,
to appear in B. L. Ioffe Festschrift, "{\it{At the Frontier of Particle
Physics/Handbook of QCD}}" M. Shifman ed., (World Scientific 2001),
and the references therein}. As this condensate possesses all the
symmetries of low energy hadronic matter, the cross-over from
one phase to the other is likely to be continuous
\ref{\cont}{T. Schafer and F. Wilczek, \prl82 (1999) 3956.}.
The second approach
\ref{\son}{D. T. Son, \pr59 (1999) 094019.},\ref{\SW}{T. Schafer and F. 
Wilczek, \pr60 (1999) 114033.},\ref{\Pisarski}{R. D. Pisarski and D. H. 
Rischke, \pr61 (2000) 051501.}, \ref{\bio}{R. D. Pisarski and D. H. 
Rischke, \pr61 (2000) 074017.},\ref{\hong}{D. K. Hong, \np582
(2000) 451.},
\ref{\ren}{W. Brown, J. T. Liu, H. C. Ren, 
\pr61 (2000) 114012.},\ref{\liu}{W. Brown, J. T. Liu and H. C.  Ren,
\pr62 (2000) 054016, \pr62 (2000) 054013.}, \ref{\myk}
{S. D. Hsu and M. Schwetz \np572 (2000) 211.} yielded as a
main result an asymptotic formula of the superconducting
energy scale in the weak coupling limit. For example, the transition
temperature with $N_c$ colors and $N_f$ flavors is given by
$$
k_BT_C=512\pi^3e^\gamma\Big({2\over N_f}\Big)^{5\over 2}
{\mu\over g^5}e^{-\sqrt{{6N_c\over N_c+1}}{\pi^2\over g}
-{\pi^2+4\over 16}(N_c-1)}
\numbereq\name{\eqvirus}
$$
with $g$ being the QCD running coupling constant
at $\mu$ and $\gamma=0.5772..$
the Euler constant. The non-BCS behavior of the exponent
stems from the long range propagation of the magnetic gluon being
exchanged and the $O(1)$ term of the exponent is due to the suppression
of the quasi-particle weight at the Fermi level, another consequence of
the long range magnetic
interaction. Other contributions which
are subleading in $g$ [\liu], including
the scheme dependence of the
ultraviolet renormalization
do not appear in (\eqvirus). Remarkably, the formula (\eqvirus),
even when is
extrapolated to the medium high chemical potential, gives rise to
a very small ratio $k_BT_C/\mu$.

In this paper, we shall derive the Ginzburg-Landau free energy functional
of color superconductivity at ultra high chemical potential. The
conventional method of derivation was developed by Gorkov \ref
{\gorkov}{L. P. Gorkov, \jetp36(9), No. 6, (1959) 1364.}, who
generalized the superconducting gap equation to a set of
equations for
ordinary and anomalous Green functions, the latter being appropriate
to an inhomogeneous system. Although this formalism
is easy to
implement in the case
of a four-fermion interaction it becomes rather cumbersome
when one uses the fundamental
QCD action with an energy dependent gap.
Furthermore the systematics is not transparent.
Instead, we shall employ the formulation 
developed by Jona-Lasinio \ref{\jona}
{G. Jona-Lasinio, \nuo26 (1975) 99.},
in which the Ginzburg-Landau free energy is identified
as the generating
functional of proper vertices. Each coefficient of the
Ginzburg-Landau free
energy is associated with a set of thermal diagrams
in the normal phase. As
we shall see the transition temperature
corresponds to the zero mode
of the inverse di-quark propagator and this
criterion is rigorously equivalent
to the one used in [\ren],[\liu], in
which the transition temperature was identified with the zero
of the Fredholm determinant of the di-quark scattering
kernel. For a
homogeneous system, the derived Ginzburg-Landau 
free energy agrees with the recent
result of Iida and Baym \ref{\Iida}{K. Iida and G. Baym, \pr63
(2001) 074018.}
to the leading order in $g$ and in the absence of
the color symmetric component of the
condensate. The color-flavor locked condensate
remains the energetically favored minimum in the bulk
near $T_C$. Including the inhomogeneity,
the Ginzburg-Landau free energy
for QCD agrees exactly with that of a BCS superconductor
upon setting the
Fermi velocity of the latter to the speed of
light, despite the non-BCS
scaling of the transition temperature (\eqvirus).

Although the one gluon exchange is repulsive in the
color symmetric channel,
a small component of the condensate in this channel
was found to be induced
at $T=0$ [\alford], \ref{\schafer}{T. Schafer, \np575 (2000) 269.},
\ref{\shovkovy}{I. A. Shovkovy and L. C. R. Wijewardhana, 
\pl470 (1999), 189.}. We shall analyze this mechanism as
$T$ approaches $T_C$
from below. We find that the magnitude of the induced color symmetric
condensate scales like the cube of that of the color antisymmetric
condensate. The contribution to the free energy at equilibrium from the
former is suppressed relative to that from the latter by a factor of the
order of $(T_C-T)\over T_C$, which is larger than the
one that was
speculated in \ref{\brook}{D. H. Rischke and R. D. Pisarski,
{\it Color Superconductivity in Cold, Dense Quark Matter},
Proceedings of the 5th Workshop on QCD, Villefrance-sur-Mer,
France, 3-7 January 2000.}.

As an important application of the Ginzburg-Landau
theory, we shall also
examine the coupling between the color-flavor
locked condensate and the
unlocked one through a gauge potential for an inhomogeneous system. 
We find the inevitable appearance of the unlocked condensate in
the core region of a vortex filament or at the border of the super 
phase while the locked condensate remains dominant in the bulk. 

This paper is organized as follows: In section 2 we review the
Jona-Lasinio method which we shall employ in this paper while
in section 3 we discuss different characheristics of the
order parameter of color superconductivity, which
corresponds to a non-zero
vacuum expectation value of a quark bilinear.
Next in section 4 we calculate the Ginzburg-Landau
free energy functional to the quadratic order in
the order parameter and in section 5 we
extend the calculation to the quartic
order. Finally in section 6 we introduce a non-zero
total momentum to the diquark condensate and the corresponding 
gauge coupling and examine the application to the realistic
case with three
colors and three flavors.

\newsection General Formulation.

In this section, we shall review
the Jona-Lasinio method [\jona], which we shall 
employ in the microscopic derivation of the Ginzburg-Landau free energy
functional of color superconductivity through out this paper.

Let ${\cal L}_E$ be the Lagrangian density which
describes a field theoretic
system in Euclidean space. The free energy of the system
at temperature $T$,
$F(T)$ can be expressed in terms of a functional integral
$$
e^{-\beta F(T)}=C\int\prod_ad\phi_ae^{-S_E(\phi)},
\numbereq\name{\eqeleu}
$$
where $\beta=1/k_BT$, $S_E(\phi)
=\int_0^\beta d\tau\int d^3\vec r{\cal L}_E
(\phi)$  and
$\phi_a$ stands for the collection of the field variables of the 
system with the subscript $a$ denoting its spacetime and internal 
coordinates.
Periodic boundary conditions (anti-periodic)
in Euclidean time $\tau$ are
imposed on the bosonic fields (fermionic).

The order parameter of the system corresponds to the nonzero expectation 
value of an operator, $\Phi_a$, which may be one
of the elementary fields of the theory or
a composite operator and realize a
nontrivial representation of the symmetry 
group of the system. To explore the stability of the system against the order 
parameter, we first trigger it with an external source, i.e.
$$
S_E(\phi)\to S_E^J(\phi)=S_E(\phi)+\sum_a(J_a^*\Phi_a+J_a\Phi_a^*).
\numbereq\name{\eqaman}
$$
The corresponding shift of the free energy, $W(J)$, is given by
$$
e^{-\beta W(J)}={\int\prod_ad\phi_ae^{-S_E^J(\phi)}\over\int\prod_ad\phi_a
e^{-S_E(\phi)}},
\numbereq\name{\eqpatsa}
$$
and can be expanded in powers of $J$ and $J^*$, i.e.
$$
W(J)=-{1\over {\beta}}\sum_{ab}{\cal D}_{ab}J_a^*J_b
+{1\over {2\beta}}\sum_{abcd}G_{abcd}J_a^*J_b^*J_cJ_d
+O(|J|^6),
\numbereq\name{\eqmauro}
$$
where each coefficient corresponds to a set of
connected diagrams that contains a fixed
number of $J$'s. In particular, ${\cal D}_{ab}$
corresponds to the propagator of $\Phi_a$. The order
parameter triggered by $J$ reads
$$
B_a\equiv<\Phi_a>=\beta{\delta W\over\delta J_a^*}=-\sum_b
{\cal D}_{ab}J_b
+\sum_{bcd}G_{abcd}J_b^*J_cJ_d+O(|J|^5).
\numbereq\name{\eqtzortzo}
$$
After performing a Legendre trasformation, 
$$
\Gamma(B)=W(J)-\sum_a(J_a^*B_a+J_aB_a^*),
\numbereq\name{\eqanato}
$$
we find that
$$
{{\delta\Gamma}\over {{\delta}B_a^*}}=-J_a.
\numbereq\name{\eqbermou}
$$
Inverting the series (\eqtzortzo), we obtain that
$$
\Gamma(B)=\sum_{ab}{\cal M}_{ab}B_a^*B_b
+{1\over 2}\sum_{abcd}{\cal G}_{abcd}B_a^*B_b^*B_cB_d+O(|B|^6)
\numbereq\name{\eqkwst}
$$
with
$$
{\cal M}_{ab}=({\cal D}^{-1})_{ab}
\numbereq\name{\eqpassi}
$$
and
$$
{\cal G}_{abcd}=\sum_{a^\prime b^\prime c^\prime d^\prime}
{\cal M}_{aa^\prime}^*{\cal M}_{bb^\prime}^*
G_{a^\prime b^\prime c^\prime d^\prime}{\cal M}_{cc^\prime}
{\cal M}_{dd^\prime}.
\numbereq\name{\eqniniad}
$$
The functional $\Gamma(B)$ is nothing but
the Ginzburg-Landau free energy
functional of the order parameter.
When the external source is removed, the 
minimization of $\Gamma$ at a nonzero $B$
signifies the condensate phase. For a second order
phase transition, the sign of the quadratic term plays a decisive role.

Since the quartic term is positive and
we are interested in the physics near $T_c$
we shall subsequently neglect higher terms. 
Few remarks are in order:

1. If the operator $\Phi_a$ is an elementary field or
a quadratic function of the elementary fields,
the manipulation leading from (\eqaman) to (\eqkwst) can be 
carried out for a free field system with
all the coefficients of (\eqkwst)
known explicitly. In this case $\Gamma$ is always positive
for a nonzero $B$
and the only minimum of $\Gamma$ occurs at $B=0$.
This is also the case with
a repulsive interaction.

2. The presence of an attractive interaction makes a
difference. Let us diagonalize
the matrix ${\cal M}$ and expand the order parameter $B$
in terms of the 
eigenmodes of ${\cal M}$. As we shall see below, in the presence 
of a Fermi sea, no matter how weak is the 
attraction, there is always a mode,
whose corresponding eigenvalue
becomes negative below a sufficiently low temperature $T_C$.
We shall refer to this mode
as the pairing mode in contrast to the others (whose
eigenvalues are positive) which we shall refer to
as non-pairing modes.
Technically, we may switch off the interaction when
dealing with the
non-pairing modes in the leading order weak coupling approximation
since the interaction is 
merely perturbative to them. The same approximation applies
to the quartic 
term of (\eqkwst), since the pairing singularity of
$G_{abcd}$ is
cancelled by the zero of ${\cal M}$'s upon amputation.

3. It is not always true that the order parameter which
minimizes $\Gamma$ 
consists of only the pairing mode. In the presence of a term
that is cubic in the pairing mode 
and linear in a nonpairing mode, an expectation value
of the nonpairing mode
will be induced upon minimization, the magnitude of
which will be proportional to
the cubic power of that of the pairing mode.
As the critical temperature 
$T_C$ is approached from below, the pairing mode condensate 
scales as $(T_C-T)^{1\over 2}$ and the induced nonpairing mode
scales as $(T_C-T)^{3\over 2}$. As we shall see below,
this is the case with the
sextet condensate in the color-flavor locked phase
of three colors and three flavors.

4. As is the case for all mean field theories
in three dimensions, the weak 
coupling calculation can not be extended to the
temperature arbitrarily close 
to the critical temperature $T_C$, since the perturbative 
expansion in the coupling constant is not uniform in $T-T_C$. 
The size of the region where our 
approximation fails, the critical region,
can be estimated according to the
Ginzburg criterion. Because of the agreement of our derived 
Ginzburg-Landau free energy with that of a BCS model, we
conclude that the width 
of the critical window is given by
$$
{|T-T_C|\over T_C}\sim\Big({k_BT_C\over \mu}\Big)^4,
\numbereq\name{\eqjava}
$$
which is exceedingly narrow.
Technically, the critical point prevents us from 
continuing $\Gamma(B)$ obtained in the normal
phase to the super phase at
$J=0$. A nonzero triggering source $J$ is
required for a smooth cross over to the
super phase. To be more specific,
the normal phase $\Gamma$ functional can
be carried over to the super phase following the path $ABCD$ in Fig.1.

In the appendix A, we shall utilize the Jona-Lasinio formulation
in order to
derive the Ginzburg-Landau free energy functional for a BCS model
as an illurstration and as comparison to the analogous
expression for color superconductivity.

\newsection The Order Parameters of QCD at High Baryon Density.

Consider now an $SU(N_c)$ color gauge theory coupled
to $N_f$ flavors of massless
quarks at temperature $T$ and chemical potential
$\mu$. The Eucliden action of the system reads
$$
S_E=\int_0^\beta d\tau
\int d^3\vec r\Big[{1\over 4}F_{\mu\nu}^lF_{\mu\nu}^l
-\bar\psi\gamma_\mu
\Big({{\partial}\over {\partial x_\mu}}-igA_\mu\Big)\psi
+\mu\psi^\dagger\psi+\cdots \Big],
\numbereq\name{\eqqcdse}
$$
where $F_{\mu\nu}^l=\partial_\mu A_\nu^l-\partial_\nu A_\mu^l+gf^{lab}
A_\mu^aA_\nu^b$, $A_\mu=A_\mu^lT^l$ with $T^l$
being the $SU(N_c)$ generators in
the fundamental representation and the dots represent
gauge fixing and ghost terms.
Since the action (\eqqcdse) is diagonal
with respect to both flavor and chirality, the
corresponding indices have been dropped in (\eqqcdse). 
Our gamma matrices are all
hermitian with $\gamma_5=\gamma_1
\gamma_2\gamma_3\gamma_4$.

The fermion field can be decomposed into eigenfunctions of 
$\gamma_5$, which in turn can be separated into 
the positive energy and negative energy parts
$$
\psi_f^c(\vec r, \tau)=\psi_{fL+}^c(\vec r, \tau)+
\psi_{fL-}^c(\vec r, \tau)
+\psi_{fR+}^c(\vec r, \tau)+\psi_{fR-}^c(\vec r, \tau),
\numbereq\name{\eqorder}
$$
where we have restored the color-flavor indices for clarity.
The Fourier expansion
of $\psi_{fL(R)\pm}^c(\vec r, \tau)$ is given by
$$
\eqalign{
\psi_{fL+}^c(\vec r, \tau)&={1\over {\sqrt{\beta\Omega}}}\sum_P
e^{i\vec p\cdot\vec r-i\nu_n\tau}a_{f,P,L}^cu_L(\vec p)\cr
\psi_{fL-}^c(\vec r, \tau)&={1\over {\sqrt{\beta\Omega}}}\sum_P
e^{i\vec p\cdot\vec r-i\nu_n\tau}b_{f,P,L}^cv_L(-\vec p)\cr
\psi_{fR+}^c(\vec r, \tau)&={1\over {\sqrt{\beta\Omega}}}\sum_P
e^{i\vec p\cdot\vec r-i\nu_n\tau}a_{f,P,R}^cu_R(\vec p)\cr
\psi_{fL-}^c(\vec r, \tau)&={1\over {\sqrt{\beta\Omega}}}\sum_P
e^{i\vec p\cdot\vec r-i\nu_n\tau}b_{f,P,R}^cv_R(-\vec p)\cr}
\numbereq\name{\eqgera}
$$
and parallel expressions in terms of $\bar a_{f,P,L}^c$,
$\bar b_{f,P,L}^c$, $\bar a_{f,P,R}^c$ and
$\bar b_{f,P,R}^c$
exist for $\bar\psi_{fL(R)\pm}^c(\vec r, \tau)$, where
$P=(\vec p,-\nu_n)$ denotes an
Euclidean four momentum with
$\nu_n=2n\pi k_BT$ ($n=\pm{1\over 2},\pm{3\over 2},
...$) the Matsubara energy. The four component spinors
$u_{L(R)}(\vec p)$ satisfy the Dirac equation
$(\gamma_4p-i\vec\gamma\cdot\vec p)u_{L(R)}(\vec p)=0$ and are
normalized according to $u_{L(R)}^\dagger(\vec p)u_{L(R)}(\vec p)=1$.
Furthermore, we have $\gamma_5u_L(\vec p)=u_L(\vec p)$ and 
$\gamma_5u_R(\vec p)=-u_R(\vec p)$. Similarly $v_{L(R)}(\vec p)$
obeys the Dirac equation and the chirality conditions.
The symbol $\Omega$
stands for the volume of the system and as $\Omega
\mapsto \infty$
$$
{1\over {\beta\Omega}}\sum_P \mapsto
{1\over {\beta}}\sum_{n}\int{{d^3{\vec p}}\over {(2{\pi})^3}}
\numbereq\name{\eqambrosios}
$$
To trigger the long range order, we add to
(\eqqcdse) a source term,
$$
\eqalign{
\Delta S &={1\over 2}\sum_{h=L,R; s=\pm}
\int_0^\beta d\tau_1\int d^3\vec r_1
\int_0^\beta d\tau_2\int d^3\vec r_2\Big[\bar{\cal J}_{f_1f_2,h,s}^{c_1c_2}
(\vec r_1,\tau_1;\vec r_2,\tau_2)\cr
&\tilde\psi_{f_1,h,s}^{c_1}(\vec r_1,\tau_1)
\sigma_2\psi_{f_2,h,s}^{c_2}(\vec r_2,\tau_2)
+ {\cal J}_{f_1f_2,h,s}^{c_1c_2}(\vec r_1,\tau_1;\vec r_2,\tau_2)
\tilde{\bar\psi}_{f_2,h,s}^{c_2}(\vec r_2,\tau_2)\sigma_2
\bar\psi_{f_1,h,s}^{c_1}(\vec r_1,\tau_1)\Big] \cr}
\numbereq\name{\eqaski}
$$
with
$$
\bar{\cal J}_{f_1f_2,h,s}^{c_1c_2}(\vec r_1,-\tau_1;\vec r_2,-\tau_2) 
={\cal J}_{f_1f_2,h,s}^{*c_1c_2}(\vec r_1,\tau_1;\vec r_2,\tau_2)
\numbereq\name{\eqconj}
$$
and
$$
{\cal J}_{f_1f_2,h,s}^{c_1c_2}(\vec r_1,\tau_1;\vec r_2,\tau_2)
={\cal J}_{f_2f_1,h,s}^{c_2c_1}(\vec r_2,\tau_2;\vec r_1,\tau_1).
\numbereq\name{\eqsymm}
$$
The condition (\eqconj) is the Wick rotation of the complex
conjugation with
a real time and the symmetry property (\eqsymm) is evident from the
anticommutation relation of the $\psi$'s and the
antisymmetry of $\sigma_2$. Both of them are
shared by the induced order parameter
$$
{\cal B}_{f_1f_2,h,s}^{c_1c_2}(\vec r_1,\tau_1;\vec r_2,\tau_2)
\equiv < \tilde\psi_{f_1,h,s}^{c_1}(\vec r_1,\tau_1)
{\sigma_2}\psi_{f_2,h,s}^{c_2}(\vec r_2,\tau_2)>.
\numbereq\name{\eqaskig}
$$
To trigger a homogeneous condensate, the source should
depend only on the relative
Euclidean coordinates, $(\vec r_1-\vec r_2,\tau_1-\tau_2)$
and its Fourier expansion
reads
$$
{\cal J}_{f_1f_2,h,s}^{c_1c_2}(\vec r_1,\tau_1;\vec r_2,\tau_2)
={1\over {\beta\Omega}}\sum_Pe^{i\vec p\cdot(\vec r_1-\vec r_2)
-i\nu_n(\tau_1-\tau_2)}
J_{f_1f_2,h,s}^{c_1c_2}(P)
\numbereq\name{\eqgeron}
$$
with $J_{f_1f_2,h,s}^{c_1c_2}(P)=J_{f_2f_1,h,s}^{c_2c_1}(-P)$.
Correspondingly,
$$
{\cal B}_{f_1f_2,h,s}^{c_1c_2}(\vec r_1,\tau_1;\vec r_2,\tau_2)
={1\over {\beta\Omega}}\sum_Pe^{i\vec p\cdot(\vec r_1-\vec r_2)
-i\nu_n(\tau_1-\tau_2)}
B_{f_1f_2,h,s}^{c_1c_2}(P)
\numbereq\name{\eqartis}
$$
with $B_{f_1f_2,h,s}^{c_1c_2}(P)=B_{f_2f_1,h,s}^{c_2c_1}(-P)$.
The triggering part of the
action (\eqaski) becomes
$$
\Delta S={\sum_{P}}^\prime [{\ov J}_{f_1f_2,h,s}^{c_1c_2}(P)
a_{f_2,-P}^{c_2}
a_{f_1,P}^{c_1}+J_{f_1f_2,h,s}^{c_1c_2}(P)\bar 
a_{f_1,P}^{c_1}\bar a_{f_2,-P}^{c_2}]
\numbereq\name{\eqbartok}
$$
with the sum over $P$ extending only to the half $\vec p$-space
and $\ov J=J^{\star}$.

As the condensate consisting of quarks of left helicity
contributes identically with the condensate of quarks
of right helicity the
total Ginzburg-Landau
free energy functioncal of the system can be written
$$
\Gamma_{total}({\cal B}_L, {\cal B}_R)={\Gamma}({\cal B}_L)
+{\Gamma}({\cal B}_R)+\cdots
\numbereq\name{\eqaleka}
$$
where the $\cdots$ indicate terms which include
both ${\cal B}_L$ and ${\cal B}_R$ and arise due
to instanton effects \ref{\alf}{M. Alford,
K. Rajagopal and F. Wilczek, \pl422 (1998) 247.}, \ref{\rapp}{R. Rapp,
T. Schafer, E. Shuryak and M. Velkovsky, \prl81 (1998) 53.}.
Such terms are exponentially small in comparison with the
first two terms of (\eqaleka) but nevertheless contribute
to fix the relative phase between ${\cal B}_L$ and ${\cal B}_R$.
Since the even parity is favored for the ground state, we
expect that
$$
{\cal B}_{f_1f_2,L,s}^{c_1c_2}(\vec r_1,\tau_1;\vec r_2,\tau_2)=
{\cal B}_{f_1f_2,R,s}^{c_1c_2}(-\vec r_1,\tau_1;-\vec r_2,\tau_2)=
{\cal B}_{f_1f_2,s}^{c_1c_2}(\vec r_1,\tau_1;\vec r_2,\tau_2)
\numbereq\name{\eqmuseum}
$$
and (\eqaleka) becomes
$$
\Gamma_{total}({\cal B}_L, {\cal B}_R)\cong 2 \Gamma({\cal B})
\numbereq\name{\eqfinal}
$$
where we have omitted the extra terms in (\eqaleka).
$\Gamma({\cal B})$ is the Ginzburg-Landau free energy
functional with a single helicity. It's quartic expansion
$$
\Gamma({\cal B})=\Gamma_2({\cal B})+\Gamma_4({\cal B})
\numbereq\name{\eqveron}
$$
will be evaluated in the subsequent sections with the helicity
subscripts L, R omitted for clarity of notation.

At high baryon density, the pairing correlation extends
hardly to the states below
the Dirac sea and we may set ${\cal J}_-=0$ with
${\cal J}_+\equiv{\cal J}$.

In this paper, we shall
consider only the $s$-wave pairing for which both
$J_{f_1f_2}^{c_1c_2}(P)$ and
$B_{f_1f_2}^{c_1c_2}(P)$ are independent of the direction of
$\vec p$. As we shall
see in the next section, the attractive force around
the Fermi surface picks the
order parameter which is even with respect to the sign
of $\nu_n$ for pairing.
Therefore
$$
J_{f_1f_2}^{c_1c_2}(P)=J_{f_1f_2}^{c_1c_2}(-P)
\numbereq\name{\eqgratif}
$$
and
$$
B_{f_1f_2}^{c_1c_2}(P)=B_{f_1f_2}^{c_1c_2}(-P).
\numbereq\name{\eqpassal}
$$
It follows from (\eqsymm) that
$$
J_{f_1f_2}^{c_1c_2}(P)=J_{f_2f_1}^{c_2c_1}(P)
\numbereq\name{\eqarchip}
$$
and
$$
B_{f_1f_2}^{c_1c_2}(P)=B_{f_2f_1}^{c_2c_1}(P).
\numbereq\name{\eqvoinov}
$$
In other words, the triggering source and the order parameter
are symmetric with respect
to simultaneous exchange of color and flavor indices.
According to the antisymmetric and symmetric representations 
of the color group, we decompose further
$$
B_{f_1f_2}^{c_1c_2}(P)=\phi_{f_1f_2}^{c_1c_2}(P)
+\chi_{f_1f_2}^{c_1c_2}(P),
\numbereq\name{\eqkokka}
$$
with 
$$
\phi_{f_1f_2}^{c_1c_2}(P)=-\phi_{f_1f_2}^{c_2c_1}(P)
=-\phi_{f_2f_1}^{c_1c_2}(P)=\phi_{f_2f_1}^{c_2c_1}(P)
$$
and
$$
\chi_{f_1f_2}^{c_1c_2}(P)=\chi_{f_1f_2}^{c_2c_1}(P)
=\chi_{f_2f_1}^{c_1c_2}(P)=\chi_{f_2f_1}^{c_2c_1}(P)
$$
with $\phi_{f_1f_2}^{c_1c_2}(P)$ pertaining to the attractive 
channel and $\chi_{f_1f_2}^{c_1c_2}(P)$ to the repulsive 
channel.
The color-flavor locked condensate for
$N_c=N_f$ corresponds, up to
color-flavor-baryon number
rotations, to
$$
B_{f_1f_2}^{c_1c_2}(P)=B_1(P)\delta_{f_1}^{c_1}\delta_{f_2}^{c_2}
+B_2(P)\delta_{f_1}^{c_2}\delta_{f_2}^{c_1},
\numbereq\name{\eqarenas}
$$
which gives rise to the minimum free energy of
a homogeneous system below $T_C$.

\newsection The Di-Quark Propagator of QCD at High Baryon Density.

The coefficient of $\Gamma_2(\cal B)$ of
(\eqkwst) is the inverse of the
di-quark propagator, which corresponds to the
thermal diagrams in Fig.2, when it is sandwiched by
two order parameters. For
total zero momentum and total zero Matsubara energy,
the propagator is specified by
the relative momentum, relative energy and color-flavor indices at both 
terminals. In writing, we have
$$
\eqalign{
<P^\prime,s_1^\prime,s_2^\prime;c_1^\prime,c_2^\prime;
f_1^\prime,f_2^\prime|{\cal D}&|P,s_1,s_2;c_1,c_2;f_1,f_2>=
\delta^{c_1^\prime c_1}\delta^{c_2^\prime c_2}\delta_{f_1^\prime f_1}
\delta_{f_2^\prime f_2}\delta_{P^\prime P}
D_{s_1^\prime s_2^\prime,s_1s_2}(P)\cr
&+\delta_{f_1^\prime f_1}\delta_{f_2^\prime f_2}\sum_{t_1^\prime t_2^\prime,
t_1 t_2} D_{s_1^\prime s_2^\prime,t_1^\prime t_2^\prime}
(P^\prime)\Gamma_{t_1^\prime t_2^\prime,t_1 t_2}^{c_1^\prime
c_2^\prime,c_1c_2}(P^\prime|P)
D_{t_1t_2, s_1s_2}(P), \cr}
\numbereq\name{\eqDS}
$$
where $D$ represents the
disconnected diagrams of the di-quark propagator, the first diagram of 
Fig. 2a, $\Gamma$ is the sum of the
proper vertices of the scattering between 
two quarks with zero total 
momentum and energy, and $P=(\vec p,-\nu_n)$
denotes an Euclidean
four-momentum. The superscripts $c_1,c_2,c_1^\prime,c_2^\prime$ 
stand for color indices, the subscripts
$f_1,f_2,f_1^\prime,f_2^\prime$
denote the flavor indices and the subscripts
$s_1,s_2,s_1^\prime,s_2^\prime$
and  $t_1,t_2,t_1^\prime,t_2^\prime$ label the
states above or below the Dirac
sea. In order to facilitate the partial wave
analysis, we found it convenient
to associate the Dirac spinors $u(\vec p)$
and $v(\vec p)$
to the vertex instead of the
propagator. Thus the vertex in (\eqDS) is of the form
$$
\Gamma_{s_1^\prime s_2^\prime,s_1 s_2}^{c_1^\prime c_2^\prime,c_1c_2}
(P^\prime|P)=\bar U_\gamma(s_1^\prime,\vec p^\prime)
\bar U_\delta(s_2^\prime,-\vec p^\prime)
\Gamma_{\gamma\delta\alpha\beta}^{c_1^\prime c_2^\prime,c_1c_2}
(P^\prime|P)U_\alpha(s_1,\vec p)U_\beta(s_2,-\vec p)
\numbereq\name{\eqvertex}
$$
with the vertex function $\Gamma_{\gamma\delta\alpha\beta}$ given by 
conventional Feynman rules and $U(s,\vec p)=u(s\vec p)$. The overall
sign of (\eqDS) is fixed such that
$D_{++++}(P) \mapsto {1\over\nu_n^2+(p-\mu)^2}$
as $g\to 0$.

The proper vertex function $\Gamma$ satisfies
the Schwinger-Dyson equation:
$$
\eqalign{
\Gamma_{s_1^\prime s_2^\prime,s_1 s_2}^{c_1^\prime
c_2^\prime,c_1c_2}(P^\prime|P)
&=\bar\Gamma_{s_1^\prime s_2^\prime,s_1 s_2}^{c_1^\prime
c_2^\prime,c_1c_2}(P^\prime|P)\cr
&+{1\over {\beta\Omega}}
\sum_{P^{\prime\prime};t_1^\prime,t_2^\prime,t_1,t_2;
c_1^{\prime\prime},c_2^{\prime\prime}}
\bar\Gamma_{s_1^\prime s_2^\prime,t_1^\prime t_2^\prime}^{c_1^\prime
c_2^\prime,c_1^{\prime\prime}c_2^{\prime\prime}}(P^\prime|P^{\prime\prime})
D_{t_1^\prime t_2^\prime t_1t_2}(P^{\prime\prime})
\Gamma_{t_1t_2,s_1 s_2}^{c_1^{\prime\prime}
c_2^{\prime\prime},c_1c_2}(P^{\prime\prime}|P)\cr}
\numbereq\name{\eqSD}
$$
with $\bar\Gamma$ standing for two particle
irreducible vertices.
In terms of matrix notations, (\eqDS) and (\eqSD) take the form
$$
{\cal D}=D-D\Gamma D
\numbereq\name{\eqmatrix}
$$
and
$$
\Gamma=\bar\Gamma+\bar\Gamma D\Gamma
\numbereq\name{\eqmatr}
$$
Combining (\eqmatrix) and (\eqmatr), we find a simple 
relationship
$$
{\cal M}={\cal D}^{-1}=D^{-1}+\bar\Gamma,
\numbereq\name{\eqmass}
$$
which generalizes the diagrammatics of a single particle 
propagator in terms of one particle irreducible
self energy function. Upon
decomposition of $\bar\Gamma$ into its symmetric
(sextet color representation
of $SU(N_c)$) and antisymmetric (anti-triplet color representation 
of $SU(N_c)$) components, 
$$
\eqalign{
\bar\Gamma_{s_1^\prime s_2^\prime,s_1 s_2}^{c_1^\prime
c_2^\prime,c_1c_2}(P^\prime|P)
&={1\over 2}(\delta^{c_1^\prime c_1}\delta^{c_2^\prime c_2} 
-\delta^{c_1^\prime c_2}\delta^{c_2^\prime c_1})
{\bar\Gamma}_{s_1^\prime s_2^\prime,s_1 s_2}^A(P^\prime|P)\cr
&+{1\over 2}(\delta^{c_1^\prime c_1}\delta^{c_2^\prime c_2} 
+\delta^{c_1^\prime c_2}\delta^{c_2^\prime c_1})
{\bar\Gamma}_{s_1^\prime s_2^\prime,s_1 s_2}^S(P^\prime|P), \cr}
\numbereq\name{\eqkare}
$$
we obtain
$$
\eqalign{
&<P^\prime,s_1^\prime,s_2^\prime;c_1^\prime,c_2^\prime;
f_1^\prime,f_2^\prime|{\cal M}|P,s_1,s_2;c_1,c_2;f_1,f_2>
={1\over 2}\delta_{f_1^\prime f_1}\delta_{f_2^\prime f_2}
\Big[(\delta^{c_1^\prime c_1}\delta^{c_2^\prime c_2}
-\delta^{c_1^\prime c_2}\delta^{c_2^\prime c_1})\cr
&<P^\prime,s_1^\prime,s_2^\prime|{\cal M}_A|P,s_1,s_2>
+(\delta^{c_1^\prime c_1}\delta^{c_2^\prime c_2} 
+\delta^{c_1^\prime c_2}\delta^{c_2^\prime c_1})
<P^\prime,s_1^\prime,s_2^\prime|{\cal M}_S|P,s_1,s_2>\Big]\cr}
\numbereq\name{\eqmassAS}
$$
with
$$
<P^\prime,s_1^\prime,s_2^\prime|{\cal M}_{A(S)}|P,s_1,s_2>
=\delta_{P^\prime P}
D^{-1}_{s_1^\prime s_2^\prime,s_1s_2}(P)
+{1\over {\beta\Omega}}
{\bar\Gamma}_{s_1^\prime s_2^\prime,s_1 s_2}^{A(S)}(P^\prime|P).
\numbereq\name{\eqASmass}
$$
To the order of the approximation made in
this work, the states below the Dirac
sea do not contribute. We shall consider only the channels
with all states above the Dirac sea
and have the subscripts $s$'s and $t$'s suppressed.  

By calculating the matrix elements of ${\cal M}$
between the order parameter
(\eqkokka), we obtain formally the quadratic term of the 
Ginzburg-Landau free energy functional with a single helicity
$$
\Gamma_2({\cal B})={1\over {2\beta}}\sum_{P^\prime,P}
[<P^\prime|{\cal M}_A|P>{\rm{Tr}}\phi^\dagger(P^\prime)
\phi(P)+<P^\prime|{\cal M}_S|P>{\rm{Tr}}\chi^\dagger(P^\prime)\chi(P)],
\numbereq\name{\eqpente}
$$
where we regard $\phi_{f_1f_2}^{c_1c_2}$
and $\chi_{f_1f_2}^{c_1c_2}$ matrix elements
of $N_{c}N_{f} \times N_{c}N_{f}$ matrices $\phi$ and $\chi$.
The symmetry factor $1\over 2$ stems from the fact that we
are only considering the condensate that consists
of quarks of left helicity.
With the HDL resummation, the irreducible vertex
for the two quark scattering 
above the Dirac sea reads
$$
\bar\Gamma_A(P^\prime|P)\simeq
-{{g^2}\over {12pp^\prime}}
\Big(1+{1\over {N_c}}\Big)\ln{1\over {|\hat\nu_n
-\hat\nu_{n^\prime}|}}+\hbox{higher partial wave components}
\numbereq\name{\eqirr}
$$
with $$\hat\nu={{g^5}\over {256\pi^4}}
\Big({{N_f}\over 2}\Big)^{5\over 2}{{\nu}\over
{\mu}}$$ and the function $D^{-1}$ for two quarks
propagating above Dirac sea
is given by
$$
D^{-1}(P)=[i\nu_n-p+\mu-\Sigma(i\nu_n,p)][-i\nu_n-p+\mu
-\Sigma(-i\nu_n,p)]
\numbereq\name{\eqinverse}
$$
with the single quark self-energy
$$
\Sigma(i\nu_n,p)\simeq -i{{g^2}\over{12\pi^2}}
{{N_{c}^2-1}\over {2N_c}}\nu_n\ln
{{4q_c^3}\over {\pi m_D^2|\nu_n|}},
\numbereq\name{\eqself}
$$
where $m_D={{N_fg^2{\mu}^2}\over {2{\pi}^2}}$ denotes
the Debye mass and $q_c$ is an
infrared cutoff ($|\nu_n|<<q_c<<\mu$). As
it was shown in [\liu], the weak coupling
approximation in (\eqirr)
and (\eqself) is sufficient to determine
the leading order of the pre-exponential factor of the
transition temperature
(\eqvirus) or the gap energy. Since the color-symmetric channel does
not carry the pairing
modes, we may set
$$
<P^\prime|{\cal M}_S|P>\simeq D_0^{-1}(P)\delta_{P^\prime P}
\numbereq\name{\eqsvertex}
$$
with $D_0(P)={1\over {\nu_n^2+(p-\mu)^2}}$ the free di-quark thermal
propagator.

The spectrum of the operator ${\cal M}_A$ is given
by the eigenvalue problem,
$$
{1\over {\beta\Omega}}\sum_{P^\prime}
<P|{\cal M}_A|P^\prime>u(P^\prime)=Eu(P)
\numbereq\name{\eqspectrum}
$$
i.e.
$$
D^{-1}(P)u(P)+{1\over {\beta\Omega}}\sum_{P^\prime}
\Gamma_A(P|P^\prime)u(P^\prime)=Eu(P)
\numbereq\name{\eqspectrum1}
$$
Introducing 
$$
f(P)\equiv-{1\over {\beta\Omega}}\sum_{P^\prime}
\Gamma_A(P|P^\prime)u(P^\prime)
\numbereq\name{\eqcoeff}
$$
we have
$$
u(P)={{f(P)}\over {D^{-1}(P)-E}}
\numbereq\name{\eqeigenf}
$$
Substituting (\eqeigenf) back to (\eqcoeff), we find that the
function $f(P)$ satisfies a homogeneous Fredholm equation
$$
f(P)={1\over {\beta\Omega}}\sum_{P^\prime}K_E(P|P^\prime)f(P^\prime)
\numbereq\name{\eqfredholm}
$$
with the kernel given by
$$
K_E(P^\prime|P)=-{{\Gamma_A(P^\prime|P)}\over {D^{-1}(P)-E}}.
\numbereq\name{\eqkernel}
$$
In order to
demonstrate the equivalence  with
the methodology used in [\ren],[\liu] to determine
the transition temperature,
we switch to the eigenvalue problem defined by the Fredholm equation
$$
h(P)={{\lambda^2}\over {\beta\Omega}}\sum_{P^\prime}
K_E(P|P^\prime)h(P^\prime)
\numbereq\name{\eqfredholma}
$$
with the eigenvalues $\lambda^2$'s functions of the
temperature, chemical
potential and $E$. The eq. (\eqfredholm) implies that
$$
\lambda^2(T,\mu,E)=1
\numbereq\name{\eqcond}
$$
which determines $E$ as function of $T$ and $\mu$.
For the pairing mode, we
expect that $E=0$ at $T=T_C$, i.e. $\lambda^2(T_C,\mu,0)=1$,
which is exactly the criterion for $T_C$ in [\liu]. As $T$ gets
sufficiently close to
$T_C$, we may write
$$
K_E(P|P^\prime)=K(P|P^\prime)+\delta K(P|P^\prime)
\numbereq\name{\eqpert}
$$
with $K(P|P^\prime)=-\Gamma_A(P|P^\prime)D(P^\prime)$ and 
$\delta K(P|P^\prime)=-\Gamma_A(P|P^\prime)D^2(P^\prime)E$.
Correspondingly,
$$
1=\lambda^{-2}(T,\mu,0)+c(T,\mu)E,
\numbereq\name{\eqeigenE}
$$
with the coefficient $c(T, \mu)$ being calculated perturbatively. The
eigenvalues of the kernel
at $E=0$ and the transition temperature have been
explored in [\ren], [\liu], and the one
that satisfies the condition $\lambda^2(T_C,\mu,0)=1$ is given, in the
vicinity of $T_C$, by
$$
{1\over {\lambda^2}}={{g^2}\over {6\pi^4}}
\Big(1+{1\over {N_c}}\Big)\ln^2{1\over
{\epsilon}}\Big[1+2{{\gamma+\ln2}\over {\ln{1\over{\epsilon}}}}
-{{g^2}\over {48\pi^4N_c}}(N_{c}^2-1)(\pi^2+4)\ln{1\over {\epsilon}}
+O(g^2)\Big],
\numbereq\name{\eqvion}
$$
where
$$
\epsilon=
\Big({{N_f}\over 2}\Big)^{{5\over 2}}{{g^5k_BT}\over {256\pi^3\mu}}
\numbereq\name{\eqaceh}
$$
and
$\ln{1\over {\epsilon}}\sim{1\over g}$, so
that the 2nd and 3rd term inside the
bracket are of the same order, $O(g)$. For $|T-T_C|<<T_C$, we have
$$
\eqalign{
1-{1\over {\lambda^2(T,\mu,0)}}&\simeq
{{g^2}\over {3\pi^4}}\Big(1+{1\over {N_c}}\Big)
\Big[\ln{1\over {\epsilon_C}}+\gamma+\ln2
-{3\over 16}(N-1)(\pi^2+4)\Big]
{{T-T_C}\over T_C} \cr
&\simeq{{g^2}\over {3\pi^4}}\Big(1+{1\over {N_c}}\Big)
\ln{1\over {\epsilon_C}}{{T-T_C}\over {T_C}}, \cr}
\numbereq\name{\eqalpha}
$$
with $\epsilon_c={\epsilon}|_{T=T_c}$.
The constant pertaining the logarithm in the braket
of (\eqalpha), which
contributes to the pre-exponential factor of the
transition temperature (\eqvirus)
is subleading in the coefficient of $T-T_c$
in comparison with $\ln{1\over {\epsilon_C}}$ and
is dropped in
the last step.

For the rest of the calculation, we shall set $T=T_C$
and maintain only the
leading order term in $g$. Under such an
approximation, the eigenfunction in
(\eqfredholma) at $E=0$ is given by [\liu]
$$
h(P)={{4\pi^2}\over {\sqrt{7\zeta(3)}}}{{k_BT_C}\over {p}}
\sin\Big({{\pi\ln{1\over{\hat\nu_n}}}\over
{2\ln{1\over {\epsilon_C}}}}\Big)
\theta(\delta-|\nu_n|)
\numbereq\name{\eqgap}
$$
with $\delta$ a cutoff energy such that
$k_BT_C<<\delta<<\mu$, and the
eigenfunction $u(P)$ in (\eqeigenf) that corresponds to the
pairing mode reads
$$
u(P)={{h(P)}\over{\nu_n^2+(p-\mu)^2}}.
\numbereq\name{\eqwavefunc}
$$
The choice of the constant in front of the sine
of (\eqgap) normalizes
$u(P)$ according to
$$
{1\over {\beta_C\Omega}}\sum_{P}|u(P)|^2=1.
\numbereq\name{\eqnorm}
$$
By sandwiching (\eqpert) between $h(P)$ and its adjoint,
$\bar h(P)\simeq{{h(n,\vec p)}\over {\nu_n^2+(p-\mu)^2}}$,
and subsequently
summing over $P$, we find that
$$
c(T_C,\mu)={{7\zeta(3)}\over {4\pi^2k_B^2T_C^2\ln{1\over{\epsilon_C}}}}.
\numbereq\name{\eqcc}
$$
It follows from (\eqeigenE), (\eqalpha) and (\eqcc) that
$$
E\simeq{{8\pi^2}\over {7\zeta(3)}}k_B^2T_C(T-T_C).
\numbereq\name{\eqeigenvalue}
$$
Note that, the summations over $n$ in (\eqnorm) and
(\eqcc) are
dominated by the terms with $|n|=O(1)$ and the energy
dependence of
$h(P)$ may be dropped to the leading order in $g$.

Upon the decomposition of $\phi$ into its pairing mode
and non-pairing modes,
$$
\phi=\phi_0+\phi^\prime,
\numbereq\name{\eqaristn}
$$
with
$$
\phi_0=\sqrt{6}\Psi u(P)
\numbereq\name{\eqreonb}
$$
and
$$
{1\over {\beta\Omega}}\sum_P\phi_0^\dagger(P)\phi^\prime(P)=0
\numbereq\name{\eqxrixto}
$$
the quadratic term of the Ginzburg-Landau
functional (\eqpente) becomes
$$
\Gamma_2({\cal B})=\Omega{{24\pi^2}\over {7\zeta(3)}}k_B^2T_C(T-T_C)
{\rm{Tr}}\Psi^\dagger\Psi
+{1\over {2\beta}}\{\sum_P[\nu_n^2+(p-\mu)^2]{\rm{Tr}}
[\phi^{\prime\dagger}(P)\phi^\prime(P)+\chi^\prime(P)\chi(P)]\},
\numbereq\name{\eqvuon}
$$
where $\Psi$ is a constant $N_{c}N_{f}\times N_{c}N_{f}$ matrix,
the function $u(P)$ is given by
(\eqwavefunc) and we have made a zero-th order approximation
(\eqsvertex), to the coefficients 
in front of the non-pairing modes.

\newsection The Quartic Terms.

Because of the elimination of the pairing singularity
in the diquark
scattering channel embedded in the diagrams ${\cal G}$ upon
amputation, we may
proceed with $g=0$. The action becomes then quadratic and
the path integral
in the presence of the triggering term can be evaluated explicitly.

Ignoring the Yang-Mills part and the fermionic
states below the Dirac sea, the
action (\eqqcdse) becomes
$$
S_E(a,\bar a)=\sum_{P}(-i\nu_n+p-\mu)\bar a_{f,P}^c
a_{f,P}^c+{\sum_{P}}^\prime [J_{f_1f_2}^{c_1c_2}(P)^* a_{f_2,-P}^{c_2}
a_{f_1,P}^{c_1}+J_{f_1f_2}^{c_1c_2}(P)\bar 
a_{f_1,P}^{c_1}\bar a_{f_2,-P}^{c_2}],
\numbereq\name{\eqfermion}
$$
where the repeated color-flavor indices are summed and the second sum,
$\sum_P^\prime$ extends only half $\vec p$-space. 
The triggering source for a given $P=(\vec p,-\nu_n)$,
$J_{f_1f_2}^{c_1c_2}(P)$ is an element of a $N_{c}N_{f}\times
N_{c}N_{f}$ matrix $J(P)$,
and the free energy shift is given by
$$
\eqalign{
W(J)&=-k_BT_C{\sum_P}^\prime\ln{{{\rm{det}}\left(\matrix{-i\nu_n+p-\mu&J(P)
\cr J^\dagger(P)&-i\nu_n-p+\mu\cr}\right)}\over {{\rm{det}}
\left(\matrix{-i\nu_n+p-\mu&0\cr0&-i\nu_n-p+\mu\cr}\right)}}\cr
&=-{{k_BT_C}\over {2}}\sum_P{{{\rm{Tr}}J^\dagger(P)J(P)}
\over {\nu_n^2+(p-\mu)^2}}
+{{k_BT_C}\over 4}\sum_P{{{\rm{Tr}}[J^\dagger(P)J(P)]^2}\over
{[\nu_n^2+(p-\mu)^2]^2}}+.... \cr}
\numbereq\name{\eqshift}
$$
which corresponds to the sum of the ring diagrams in Fig. 3.
Note the symmetry factor $1\over 2$ of each term because of the
single helicity pairing.
Through a Legendre transformation, we obtain the quartic
term of the Ginzburg-
Landau free energy functional with
a single helicity in terms of the order parameters
$$
\Gamma_4({\cal B})={{k_BT_C}\over 4}\sum_P[\nu_n^2+(p-\mu)^2]^2{\rm{Tr}}
[B^\dagger(P)B(P)]^2,
\numbereq\name{\eqglqurtic}
$$
the coefficient of which is valid to the leading order
of the coupling constant
$g$. Upon the decomposition of the order
parameter $B$ into its color
antisymmetric component and color symmetric
one, eq.(\eqkokka) and further 
decomposition of the color antisymmetric component into
its pairing mode and
non-pairing modes, (\eqaristn), we arrive at
$$
\eqalign{
\Gamma_4({\cal B}) &={{k_BT_C}\over 4}\sum_P[\nu_n^2+(p-\mu)^2]^2{\rm{Tr}}\{
[\phi_0^\dagger(P)\phi_0(P)]^2 \cr
&+2\phi_0^\dagger(P)\phi_0(P)[\phi_0^\dagger(P)\chi(P)
+\chi^\dagger(P)\phi_0(P)] \cr
&+2\phi_0^\dagger(P)\phi_0(P)[\phi_0^\dagger(P)\phi^\prime(P)
+\phi^{\prime\dagger}(P)\phi_0(P)]+...\} \cr}
\numbereq\name{\eqphichi}
$$
where $\cdots$ indicate terms of higher powers
of $\chi$ and $\phi^\prime$.
We would like to emphasize here that the second and third terms of
the right hand side 
of (\eqphichi) are in general nonzero. Indeed, for the color flavor 
locked condensate (\eqarenas), at $N_c=N_f=N$
$$
\eqalign{
\phi_{f_1f_2}^{c_1c_2}(P)&={1\over 2}[B_1(P)-B_2(P)]
(\delta_{f_1}^{c_1}\delta_{f_2}^{c_2}
-\delta_{f_1}^{c_2}\delta_{f_2}^{c_1})\cr
\chi_{f_1f_2}^{c_1c_2}(P)&={1\over 2}[B_1(P)+B_2(P)]
(\delta_{f_1}^{c_1}\delta_{f_2}^{c_2}
+\delta_{f_1}^{c_2}\delta_{f_2}^{c_1}), \cr}
\numbereq\name{\eqmakiko}
$$
we find that
$$
\eqalign{
&{\rm{Tr}}\phi^\dagger(P)\phi(P)\phi^\dagger(P)\chi(P)=\cr
&={{N(N^2-1)(N-2)}\over {16}}|B_1(P)-B_2(P)|^2
[B_1(P)-B_2(P)]^*[B_1(P)+B_2(P)]. \cr}
\numbereq\name{\eqmaiko}
$$
which is nonzero for $N>2$.
By combining (\eqvuon) and (\eqphichi), we derive an expression
for the Ginzburg-Landau free energy
functional for a homogeneous system,
$$
\eqalign{
\Gamma({\cal B}) &={{k_BT_C}\over 2}\sum_{P^\prime P}<P^\prime|{\cal M}_A|P>
{\rm{Tr}}\phi_0^\dagger(P^\prime)\phi_0(P)\cr
&+{{k_BT_C}\over 2}\sum_P[\nu_n^2+(p-\mu)^2]
{\rm{Tr}}[\phi^{\prime\dagger}(P)\phi^\prime(P)+\chi^\dagger(P)\chi(P)]\cr
&+{{k_BT_C}\over 4}\sum_P(\nu_n^2+(p-\mu)^2]^2{\rm{Tr}}\{
[\phi_0^\dagger(P)\phi_0(P)]^2\cr
&+2\phi_0^\dagger(P)\phi_0(P)[\phi_0^\dagger(P)\chi(P)
+\chi^\dagger(P)\phi_0(P)]\cr
&+2\phi_0^\dagger(P)\phi_0(P)[\phi_0^\dagger(P)\phi^\prime(P)
+\phi^{\prime\dagger}(P)\phi_0(P)]+...\}, \cr}
\numbereq\name{\eqkamaku}
$$
where a zeroth order approximation in $g$ has been
made for the coefficient
in front of the non-pairing modes of the quadratic terms.
Minimization with 
respect to $\chi$ yields
$$
\eqalign{
\chi_{f_1f_2}^{c_1c_2}(P) &=-{1\over 4}
[\nu_n^2+(p-\mu)^2]\{[\phi_0(P)
\phi_0^\dagger(P)\phi_0(P)]_{f_1f_2}^{c_1c_2}+[\phi_0(P)\phi_0^\dagger(P)
\phi_0(P)]_{f_1f_2}^{c_2c_1}\cr
&+[\phi_0(P)\phi_0^\dagger(P)\phi_0(P)]_{f_2f_1}^{c_1c_2}
+[\phi_0(P)\phi_0^\dagger(P)\phi_0(P)]_{f_2f_1}^{c_2c_1}\} \cr}
\numbereq\name{\eqsextet}
$$
and its contribution to the free energy is of
the order of $|\phi_0|^6$. Below
the transition temperature, $\phi_0(P)\sim (T_C-T)^{1\over 2}$ for 
$T_C-T<<T_C$ and the bulk condensate energy $\sim (T_C-T)^2$.
We also notice the difference from the $T=0$ case, where
the symmetric condensate is suppressed relatively to the
antisymmetric one by a factor of order $g$ [\schafer].
On the other 
hand, the induced color-symmetric
condensate, $\chi\sim (T_C-T)^{3\over 2}$
and its contribution to the free energy $\sim (T_C-T)^3$.
The same is true for 
the non-pairing modes in the color-antisymmetric
component, $\phi^\prime(P)$. Maintaining only
the pairing mode, we may write
$$
\phi(P)\simeq \phi_0(P)=\sqrt{6}\Psi u(P)
\numbereq\name{\eqtwelve}
$$
where $u(P)$ is the eigenfunction of the
pairing mode, (\eqwavefunc), and 
$\Psi$ is a constant $N_{c}N_{f}\times N_{c}N_{f}$ matrix with the same
symmetry of the indices as $\phi(P)$.

It follows from the expression (\eqwavefunc)
for $u(P)$, (\eqgap)
for $h(P)$ and the eigenvalue (\eqeigenvalue) that
$$
\Gamma({\cal B})=\Omega\Big[{a\over 2}{\rm{Tr}}\Psi^\dagger\Psi
+{b\over 4}{\rm{Tr}}(\Psi^\dagger\Psi)^2\Big].
\numbereq\name{\eqarkoan}
$$
with
$$
a={{48\pi^2}\over {7\zeta(3)}}k_B^2T_C(T-T_C)
\numbereq\name{\eqarenon}
$$
and
$$
b={{576\pi^4}\over {7\zeta(3)}}\Big({{k_BT_C}\over {\mu}}\Big)^2.
\numbereq\name{\eqverizon}
$$
Including both helicities, we obtain the total
Ginzburg-Landau free energy of a homogeneous system
with an even parity for the ground state
$$
\Gamma_{total}({\cal B}_L, {\cal B}_R)=
\Omega\Big[a{\rm{Tr}}\Psi^{\dagger}\Psi
+{b\over 2}{\rm{Tr}}(\Psi^{\dagger}\Psi)^2\Big].
\numbereq\name{\eqargyridis}
$$
We conclude this section with some details for
the realistic case of $N_c=N_f=3$.
With even parity the $9\times 9$ matrix $\Psi$ may be related to
a $3\times 3$ matrix via
$$
\Psi_{f_1f_2}^{c_1c_2}
=\epsilon^{c_1c_2c}\epsilon_{f_1f_2f}\Phi_f^c.
\numbereq\name{\eqwer}
$$
The trace of $\Psi$'s may be related to the trace 
of $\Phi$'s. Explicitly,
$$
\eqalign{
{\rm{Tr}}\Psi^\dagger\Psi &= 4{\rm{tr}}\Phi^\dagger\Phi \cr
{\rm{Tr}}(\Psi^\dagger\Psi)^2 &= 2[({\rm{tr}}\Phi^\dagger\Phi)^2+
{\rm{tr}}(\Phi^\dagger\Phi)^2]. \cr}
\numbereq\name{\eqvopil}
$$
Consequently, the Ginzburg-Landau free energy reduces to
$$
\Gamma_{total}=\Omega\{4a{\rm{tr}}\Phi^\dagger\Phi+b[({\rm{tr}}
\Phi^\dagger\Phi)^2+{\rm{tr}}(\Phi^\dagger\Phi)^2]\}
\numbereq\name{\eqflower}
$$
where $a$ and $b$ are given in (\eqarenon) and (\eqverizon)
at weak coupling
with $a<0$ for $T<T_C$ and $b>0$.
The minimum free energy below $T_C$ corresponds to
$$
\Phi^\dagger\Phi={{|a|}\over {2b}},
\numbereq\name{\eqbuc}
$$
i.e. $\Phi$ is proportional to an unitary
matrix. By a flavor and baryon number
rotation, it can always be brought into an unit matrix,
which leads to the standard
expression of the color-flavor locked condensate with
$$
\Psi_{f_1f_2}^{c_1c_2}
=\sqrt{{|a|}\over {2b}}(\delta_{f_1}^{c_1}\delta_{f_2}^{c_2}
-\delta_{f_1}^{c_2}\delta_{f_2}^{c_1}).
\numbereq\name{\eqenergy}
$$
The corresponding condensate energy is 
$$
(\Gamma_{total})_{\rm min}=-6{{a^2}\over {b}}\Omega
=-{{12\mu^2}\over {7\zeta(3)}}k_B^2T_C^2
\Big({{T_C-T}\over {T_C}}\Big)^2\Omega.
\numbereq\name{\eqmin}
$$
We have thus proven that the color-flavor locked
condensate remains energetically favored near $T_c$.
Using (\eqsextet) and (\eqtwelve), the induced sextet
order parameter reads
$$
\chi_{f_1f_2}^{c_1c_2}(P)=-3\sqrt{3}\Big({{|a|}\over {b}}
\Big)^{3\over 2}
[\nu_n^2+(p-\mu)^2]u^3(P)
(\delta_{f_1}^{c_1}\delta_{f_2}^{c_2}+\delta_{f_1}^{c_2}\delta_{f_2}^{c_1}).
\numbereq\name{\eqviacom}
$$
and it contributes to the condensate energy a term
$$
\Delta\Gamma_{\rm{min}}=
-{{558\zeta(5)}\over {343\zeta^3(3)}}\mu^2k_B^2T_C^2
\Big({{T_C-T}\over {T_C}}\Big)^3,
\numbereq\name{\eqxion}
$$
which is smaller than (\eqmin) by a factor ${(T_C-T)\over T_C}$.
The existence of the sextet order parameter is evident
from the composition rule
${\bf 3}\times{\bf \bar 3}\times{\bf \bar 3}
={\bf \bar 3}+{\bf \bar 3}+{\bf 6}
+{\bf 15}$ of the color $SU(3)$ representations and
its conjugate for the flavor
$SU(3)$ representations.
The sextet correction (\eqxion) may be compared with the
correction due to the constraint
of color neutrality discussed in [\Iida].
The latter is suppressed by a factor of
$({{k_BT_c}\over {\mu}})^2\ln{{\mu}\over {k_BT_c}}$
relative to (\eqmin). If ${{T_c-T}\over T}\ge
({{k_BT_c}\over {\mu}})^2\ln{{\mu}\over {k_BT_c}}$,
the sextet correction will be larger in magnitude.

\newsection Kinetic Energy and Gauge Coupling.

An important application of the Ginzburg-Landau theory
lies in systems which include inhomogeneous condensates, such
as domain walls and vortex filaments. To incorporate the
spatial variation of the condensate, a nonzero total momentum
is introduced to the di-quark propagator. As we shall see
below, the characteristic momentum corresponds to the inverse of the
coherence length, which is much longer than the thermal
wavelength of the system and diverges at $T_c$.
Therefore, the momentum dependence
of the di-quark propagator will be treated perturbatively
while the dependence of the quartic coefficient
will enter only through
the constraint of the overall momentum conservation.

A non-zero total momentum $\vec k$ will modify both the disconnected
and the connected part of the di-quark propagator.
The spectral problem (\eqspectrum1) is replaced by
$$
D_{\vec k}^{-1}(P)u_{\vec k}(P)+{1\over {\beta\Omega}}\sum_{P^\prime}
\Gamma_{A, \vec k}(P|P^\prime)u_{\vec k}(P^\prime)=Eu_{\vec k}(P)
\numbereq\name{\eqspectrum2}
$$
where we have appended the subscript $\vec k$ to indicate the
dependence on the total momentum while the relative $4$-Euclidean momentum
remains denoted by $P$. Thus the $4$-Euclidean momentum
of each quark in a pair is $({{\vec k}\over 2}\pm {\vec p},
\mp {\nu}_n)$. The homogeneous case considered in section $3$
corresponds to the special case $\vec k=0$. As it was shown
in section $3$, the quark self-energy may be omitted to the
leading order of the interaction, except in the
determination of the transition temperature,
$D_{\vec k}^{-1}(P)\simeq (i{\nu}_n-p_{+}+\mu)(-i{\nu}_n-p_{-}+\mu)$
with $p_{\pm}=|{{\vec k}\over 2}\pm {\vec p}|$. The
Fredholm equation (\eqfredholma) becomes
$$
h_{\vec k}(P)={{\lambda^2}\over {\beta\Omega}}\sum_{P^\prime}
K_{E, \vec k}(P|P^\prime)h_{\vec k}(P^\prime)
\numbereq\name{\eqfredholm2}
$$
with
$$
K_{E, \vec k}(P^\prime|P)=
-{{\bar\Gamma_{A,\vec k}(P^\prime|P)}\over {(i{\nu}_n-p_{+}+\mu)
(-i{\nu}_n-p_{-}+\mu)-E}}.
\numbereq\name{\eqkernel1}
$$
The expansion (\eqpert) is supplemented by more terms
$$
K_{E, \vec k}(P|P^\prime)=K(P|P^\prime)+\delta K(P|P^\prime)
+\delta^\prime K(P|P^\prime)+{\delta}^{\prime\prime}K(P|P^\prime)
\numbereq\name{\eqpertq}
$$
with
$$
\eqalign{
\delta^\prime K(P|P^\prime)
=-K(P|P^\prime)\Big[ -{{i{\nu}_n}\over {{\nu}_{n}^2+
(p-\mu)^2}}{\hat p}\cdot{\vec k}&-{1\over 4}{3\nu_n^2-(p-\mu)^2\over
{({\nu}_{n}^2+(p-\mu)^2)^2}}({{\hat p}\cdot{\vec k}})^2\cr
&-{{(p-\mu)}\over {4p}}
{{{\vec k}^2-({\vec k}{\hat p})^2}\over {{\nu}_{n}^2+
(p-\mu)^2}}\Big] \cr}
\numbereq\name{\eqareano}
$$
while the term ${\delta}^{\prime\prime}K(P|P^\prime)$ arises
entirely due to the momentum correction of the vertex
function.
The corresponding eigenvalue condition (\eqeigenE)
is replaced by
$$
1=\lambda^{-2}(T,\mu,0)+c(T,\mu)E+c^{\prime}(T,\mu){\vec k}^2
\numbereq\name{\eqeigenA}
$$
where $c^{\prime}(T,\mu)$ is computed perturbatively, like
$c(T,\mu)$ of (\eqcc). The first order perturbation yields
the result
$$
c^{\prime}(T_C,\mu)
=-{{7\zeta(3)}\over {24\pi^2k_B^2T_C^2\ln{1\over{\epsilon_C}}}}
=-{1\over 6}c(T_C,\mu).
\numbereq\name{\eqcc1}
$$
This contribution arises exclusively from the second term inside the
bracket of (\eqareano), while the third term contributes
a term smaller by a factor whose order is less than
$({{k_BT_c}\over {\mu}})^2\ln{{\mu}\over {k_BT_c}}$.
The second order perturbation of the first term inside
the bracket of (\eqareano) contributes to $c^{\prime}(T,\mu)$
a term that is smaller than (\eqcc1) by a
factor of order $g$.
Finally lets comment that the
contribution of the term ${\delta}^{\prime\prime}K(P|P^\prime)$
is analyzed in Appendix B, where we show
that its contribution to $c^{\prime}(T_c, \mu)$ is
comparable with that of the third term inside the bracket of
(\eqareano) and therefore can be ignored.
Finally, the eigenvalue of the
pairing mode with a small total momentum becomes
$$
E\simeq{{8\pi^2}\over {7\zeta(3)}}k_B^2T_C(T-T_C)
+{1\over 6}{\vec k}^2.
\numbereq\name{\eqeigenvalue2}
$$
Therefore, the typical momentum of the condensate is $\sim k_BT_C
\sqrt{{T_C-T\over T_C}}$, whose inverse corresponds to the
coherence length.
For a condensate consisting of superpositions of different
$\vec k$'s, i.e.
$$
B(P)={\sqrt 6}\sum_{\vec k}\Psi_{\vec k}u_{\vec k}(P)
\numbereq\name{\eqareav}
$$
the Ginzburg-Landau functional becomes
$$
\Gamma={1\over 2}\sum_{\vec k}({\vec k}^2+a){\rm{Tr}}
\Psi_{\vec k}^{\dag}\Psi_{\vec k}+{1\over 4}b
\sum_{{\vec{k_1}}{\vec{k_2}}{\vec{k_1}'}{\vec{k_2}'}}
{\rm{Tr}}\Psi_{\vec{k_1}'}^{\dag}\Psi_{\vec{k_1}}
\Psi_{\vec{k_2}'}^{\dag}\Psi_{\vec{k_2}}{\delta}_{{\vec{k_1}}
+{\vec{k_2}},{\vec{k_1}'}
+{\vec{k_2}'}}
\numbereq\name{\eqland}
$$
By transforming the condensate from the momentum
space into the coordinate space via
$$
\Psi({\vec r})={1\over {{\sqrt{\Omega}}}}\sum_{\vec k}
\Psi_{\vec k}e^{i{\vec k}{\vec r}}
\numbereq\name{\eqginz}
$$
and including both helicities with an even parity ground state
we obtain the conventional form of the Ginzburg-Landau
free energy
$$
\Gamma_{total}=\int d^3{\vec r}\Big[{\rm{Tr}}{\vec{\nabla}}
\Psi^{\dag}{\vec{\nabla}}\Psi
+a{\rm{Tr}}\Psi^{\dag}\Psi+{1\over 2}b{\rm{Tr}}
(\Psi^{\dag}\Psi)^2 \Big].
\numbereq\name{\eqrsea}
$$
It is instuctive to compare the Ginzburg-Landau free
energy (\eqrsea) with that of the ordinary BCS model.
On setting $v_F=1$ for the latter and scaling $\Psi_{\vec k}
\mapsto {\sqrt{2k_F}}\Psi_{\vec k}$, the free energy
as a polynomial in $\Psi_{\vec k}$ contains identical
coefficients as (\eqarenas).

It remains to couple the order parameter to the classical
Yang-Mills color gauge field and to the electromagnetic field.
While the coupling structure is dictated by diagrams, we shall
follow a different route by replacing the gradient in (\eqrsea)
with the gauge covariant derivative. As the final result ought
to be gauge invariant, such an approach is unambigious.
Let ${\vec A}={\vec A}^l T^l$ denote the classical vector
potential of the $SU(N_c)$ color gauge field and ${\vec{\cal A}}$
that of the electromagnetic field, of which the associated $U(1)$
rotation is a subgroup of the flavor $SU(N_f)$ group. The gauge
covariant derivative acting on a single quark field is
$$
({\vec D}{\psi}^{c})_{f}={\vec\nabla}\psi^{c}_{f}-ig{\vec A}^{cc'}
\psi^{c'}_f-ieq_{f}\psi^{c'}_f
\numbereq\name{\eqnata}
$$
with $q_f$ being the charge number of the $f$-th flavor. From here
on, only the repeated color indices in (\eqnata) are
summed. Since the order parameter behaves like a direct product of two
quark fields, the gauge covariant derivative of
$\Psi$ reads
$$
({\vec D}{\Psi})^{c_1c_2}_{f_1f_2}={\vec\nabla}
\Psi^{c_1c_2}_{f_1f_2}-ig{\vec A}^{c_1c'}
\Psi^{c'c_2}_{f_1f_2}-ig{\vec A}^{c_2c'}
\Psi^{c_1c'}_{f_1f_2}-ie(q_{f_1}+q_{f_2})
{\vec{\cal A}}\Psi^{c_1c_2}_{f_1f_2}
\numbereq\name{\eqnatali}
$$
By replacing ${\vec{\nabla}}\Psi$ of (\eqrsea) with ${\vec D}\Psi$ and
including the energy of the pure Yang-Mills part, we obtain
the gauge invariant Ginzburg-Landau free energy functional
of an inhomogeneous superconductor
$$
\Gamma_{total}=\int d^3{\vec r}\Big[{1\over 2}F_{ij}^lF_{ij}^l
+{1\over 2}({\vec{\nabla}}\times{\vec{\cal A}})^2+
{\rm{Tr}}({\vec D}\Psi)^{\dag}({\vec D}\Psi)
+a{\rm{Tr}}\Psi^{\dag}\Psi+{1\over 2}b{\rm{Tr}}
(\Psi^{\dag}\Psi)^2 \Big]
\numbereq\name{\eqrseaa}
$$
It is easy to verify that the free energy functional is invariant under the 
local gauge transformations
$$
\eqalign{
\Psi^{c_1c_2}_{f_1f_2} &\mapsto u^{c_1c_{1}^{\prime}}
u^{c_2c_{2}^{\prime}}\Psi^{c_{1}^{\prime}c_{2}^{\prime}}_{f_1f_2}
e^{-i(q_{f_1}+q_{f_2}){\alpha}}\cr
{\vec A} &\mapsto u{\vec A}u^{\dagger}
+{i\over g}u{\vec\nabla}u^{\dagger}\cr
{\vec{\cal A}} &\mapsto {\vec{\cal A}}
-{1\over e}{\vec\nabla}{\alpha} \cr}
\numbereq\name{\eqserios}
$$
The equation of motion follows from the
free energy functional by applying the variational principle.

Finally, let us explore some details for the realistic
case of $N_c=N_f=3$. It follows from (\eqnata) that
$$
({\vec D}{\Psi}^{c_1 c_2})_{f_1 f_2}
=\epsilon^{c_1 c_2 c}\epsilon_{f_1 f_2 f}({\vec D}\Phi)^{c}_{f}
\numbereq\name{\eqnatal}
$$
with
$$
({\vec D}\Phi)^{c}_{f}={\vec\nabla}\Phi^{c}_{f}-ig{\vec{\ov A}}^{cc'}
\Phi^{c'}_f-ieQ_f{\vec{\cal A}}\Phi^{c}_f
\numbereq\name{\eqmnata}
$$
where ${\vec{\ov A}}={\vec A}^l{\ov T}^l$ with ${\ov T}^l=
-T^{l*}$ being the generator of the ${\bf {\ov 3}}$ representation, and
$Q_f=q_{f_1}+q_{f_2}$ and $ff_1f_2$ represent a cyclic
permutation of $1, 2, 3$. Arranging the flavor index in the conventional
order of $u, d, s$ we find the diagonal electric charge matrix
$$
Q=diag ({2\over 3}, -{1\over 3}, -{1\over 3})=
{2\over {\sqrt 3}}{\ov T}^8.
\numbereq\name{\eqoktw}
$$
Here we have adapted an expression of $T^l$ that differs
from the standard one by a cyclic permutation of rows (columns),
in which ${\ov T}^8={1\over {2\sqrt 3}}diag (2, -1, -1)$.
The equation (\eqmnata) takes the matrix form
$$
{\vec D}\Phi={\vec\nabla}\Phi-ig{\vec{\ov A}}
\Phi-ieq{\vec{\cal A}}\Phi{\ov T}^8
\numbereq\name{\eqmnatata}
$$
with $q=-{2\over {\sqrt 3}}e$. On noting that
$\Phi=\Phi_0+\bar T^l{\Phi}_l$, we have the decomposition
$$
{\vec D}\Phi=({\vec D}\Phi)_0+{\ov T}^l{{\vec D}\Phi}_l
\numbereq\name{\eqcionb}
$$
with $({\vec D}\Phi)_0={1\over 3}{\rm{tr}}({\vec D}\Phi)$
and ${{\vec D}\Phi}_l=2{\rm{tr}}{\ov T}^l{{\vec D}\Phi}$,
and its contribution to the free energy (\eqrseaa)
$$
{\rm{tr}}({\vec D}\Phi)^{\dag}({\vec D}\Phi)=
3({\vec D}\Phi)^{*}_0({\vec D}\Phi)_0+{1\over 2}
({\vec D}\Phi)^{*}_l({\vec D}\Phi)_l
\numbereq\name{\eqbios}
$$
Now we divide the field variables into two groups, the
first group contains $\Phi_0, \Phi_8, {\vec A}_8, {\vec{\cal A}}$,
while the second one  includes $\Phi_l, {\vec A}^l$ with $l\ne 0, 8$.
A closer inspection of the structure of
${\vec D}\Phi$ reveals the absence terms in the free energy
(\eqrseaa) that are linear in the variables of the
second group. Therefore a solution of the equation of motion
exists in which only the fields of the first group acquire
non-zero values. In
what follows, we shall focus on this special case.

By setting $\Phi_l={\vec A}^l=0$ for $l \ne 0, 8$
and transforming the
remaining variables
$$
\eqalign{
\phi&=\Phi_0+{1\over {\sqrt 3}}\Phi^8, \qquad
\chi={\sqrt 2}(\Phi_0-{1\over {2\sqrt 3}}\Phi^8)\cr
{\vec V}&={\vec A}^8{\cos\theta}+{\vec{\cal A}}{\sin{\theta}}
\qquad {\vec{\cal V}}=-{\vec A}^8{\sin\theta}
+{\vec{\cal A}}{\cos{\theta}}, \cr}
\numbereq\name{\eqwasin}
$$
with $\tan\theta=-{{2e}\over {{\sqrt 3}g}}$ we arrive
at the following expression for the Ginzburg-Landau
free energy functional (\eqrseaa)
$$
\eqalign{
\Gamma_{total}=\int d^3{\vec r} \Big ( &{1\over 2}({\vec\nabla}\times
{\vec{\cal V}})^2+{1\over 2}({\vec\nabla}\times
{\vec V})^2+|({\vec\nabla}-2i{\lambda}{\vec V})\phi|^2
+|({\vec\nabla}+i{\lambda}{\vec V})\chi|^2 \cr
&+4a(|{\phi}|^2+|{\chi}|^2)+b[(|{\phi}|^2+|{\chi}|^2)^2
+|{\phi}|^4+{1\over 2}|{\chi}|^4] \Big ), \cr}
\numbereq\name{\eqgrop}
$$
with $\lambda={1\over {2{\sqrt 3}}}{\sqrt{g^2+{4\over 3}e^2}}$.
The transformation
of the gauge potentials in (\eqwasin) is nothing
but the transformation performed in [\alford], \ref{\jur}
{M. Alford, J. Berges and K. Rajagopal, \np571 (2000) 269.},
\ref{\mit}{K. Rajagopal
and F. Wilczek \prl86 (2001) 3492.} that selects
out the unbroken $U(1)$ gauge potential $\vec V$ in the
color-flavor locked phase. The $\vec{\cal V}$ field  remains
decoupled from the condensate in the presence of
a nonzero $\Phi^8$ and
can be set to zero as a special solution. The condition
for the color-flavor locking (that the $3\times 3$ matrix is
proportional to an unitary matrix),
$$
|\chi|^2=2|\phi|^2
\numbereq\name{\eqmuon}
$$
which corresponds to the free energy minimum in the bulk,
cannot be imposed everywhere because of  the nonzero values of
${\vec\nabla}\phi$, ${\vec\nabla}\chi$ and $\vec V$, as
it is evident from the example of a vortex filament
and a domain wall which are analyzed in the
appendix C. In another words, in the presence
of an inhomogeneity and of a non zero gauge potential, the unlocked
condensate, which is an octet, under a simultaneous color-flavor
rotation mixes with the locked condensate. This situation
may be compared with the $s-d$ wave mixture in the core
region of a vortex filament of a cuprite superconductor \ref{\Ting}
{Y. Ren, J. H. Xu and C. S. Ting, \prl74 (1995), 3680.}.

\newsection Concluding Remarks.

In this work, we have derived the Ginzburg-Landau free energy functional
of the color superconductivity from the fundamental QCD action by means
of normal phase thermal diagrams. We have also addressed the issue of the
induced color symmetric component of the condensate near $T_C$ and the 
interplay between the color-flavor locked condensate and the unlocked one 
in the presence of a gauge potential. Though the expressions 
of the Ginzburg-Landau coefficients were derived in the weak coupling
approximation, which is valid in the ultra-high baryon
density regime, we expect that many 
qualitative features, such as the color-flavor locking, the mixture of 
the color-symmetric condensate and the coupling among the different 
types of condensates and the gauge potential, to survive in the
regime of realistic baryon density, e.g., the baryon density achieved 
at RHIC or that inside of a neutron star.

As the chemical potential is lowered, higher order corrections become 
more important. There are two types of higher order terms, which 
constitute a hierarchy of smallness, those which
result from higher powers of the
coupling constant $g$ and those
that arise from the curvature of the density of states 
at the Fermi level. It follows from (1) that $g\sim\ln^{-1}\mu/k_BT_C$,
while the latter corrections come in powers of $(k_BT_C/\mu)^2$. 
Therefore the corrections due to 
the higher order thermal diagrams are far
more important. In view of the $2\pi$ suppression of the integral over 
loop momenta inside higher order diagrams (before the combinatorial factor
dominating the magnitude), the numerical
values of the Ginzburg-Landau coefficients
we derived may still serve as
a reasonable approximation for a realistic
high baryon density.

\noindent
\newsection Acknowlegments.

We would like like to thank D. T. Son for raising the issue of 
Ginzburg-Landau theory which motivated this reasearch. This work 
is supported in part by US Department of Energy, contract
number DE-FG02-91ER40651-TASKB.

\newsection Appendix A.

In this appendix, we shall illustrate our
methodology by rederiving the
Ginzburg-Landau free energy for the BCS model
of the system of nonrelativistic
electrons. The model Hamiltonian is given by
$$
H=\sum_{\vec ps}\xi_{\vec p}a_{\vec ps}^\dagger a_{\vec ps}
-{g\over\Omega}
\sum_{\vec p,\vec p^\prime}\theta(\omega_D-|\xi_{\vec p}|)
\theta(\omega_D-|\xi_{\vec p^\prime}|)a_{\vec p\uparrow}^\dagger
a_{-\vec p\downarrow}^\dagger a_{-\vec p^\prime\downarrow}
a_{\vec p^\prime\uparrow},
\numbereq\name{\eqbcs}
$$
where $\xi_{\vec p}={p^2\over 2m}-\mu$,
$m$ the electron mass and $\mu$
the chemical potential. The subscript $s$ labels
the spin state, $g>0$ is the
pairing coupling and $\omega_D$ denotes
the Debye frequency. The path integral
expression of the free energy at a temperature $T$ reads
$$
\exp\Big(-{F\over k_BT}\Big)=C\int\prod_{\vec ps}da_{\vec ps}
d\bar a_{\vec ps}e^{-S_E(a,\bar a)},
\numbereq\name{\eqvzoi}
$$
where the Euclidean action
$$
\eqalign{
S_E(a,\bar a)&=\sum_{Ps}(-i\nu_n+\xi_{\vec p})\bar a_{Ps} a_{Ps}\cr
&-{g\over\beta\Omega}\sum_{P_+P_-,P_+^\prime,P_-^\prime}\theta(\omega_D
-|\xi_{\vec p}|)\theta(\omega_D-|\xi_{\vec p^\prime}|)
a_{P_+\uparrow}^\dagger a_{P_-\downarrow}^\dagger
a_{P_-^\prime\downarrow} a_{P_+^\prime\uparrow}\delta_{n_++n_-,
n_+^\prime +n_-^\prime}, \cr}
\numbereq\name{\eqion}
$$
where $P=(\vec p,-\nu_n)$ is the Euclidean four momentum and
$P_\pm=(\pm\vec p, \mp\nu)$. The Matsubara
energy $\nu_n=2\pi n k_BT$ with
$n=\pm {1\over 2},\pm{3\over 2},....$.
Following the steps outlined in section 2, we add to $S_E$ a
triggering term
$$
{\Delta}S=\sum_P\theta(\omega_D-|\xi_{\vec p}|)
[J^*(P)a_{-P\downarrow}a_{P\uparrow}
+J(P)a_{P\uparrow}^\dagger a_{-P\downarrow}^\dagger],
\numbereq\name{\eqtrigger}
$$
which gives rise to an increment $W(J)$
of the free energy and induces an order
parameter
$$
B(P)={\delta W\over\delta J^*(P)}
\numbereq\name{\eqCooper}
$$
The Legendre transformation of $W(J)$,
$$
\Gamma(B)=\Gamma_2+\Gamma_4+...
\numbereq\name{\eqzizi}
$$
is the Ginzburg-Landau functional and
determines the order parameter at
equilibrium when the triggering term (\eqtrigger)
is removed, with
$$
\Gamma_2
=\sum_{P,P^\prime}B^*(P)<P|{\cal M}|P^\prime>B(P^\prime)
\numbereq\name{\eqmass}
$$
and
$$
\Gamma_4={1\over 2}
\sum_{P_1,P_2,P_1^\prime,P_2^\prime}B^{*}(P_1)B^{*}(P_2)
<P_1P_2|{\cal G}|P_1^\prime P_2^\prime>
B(P_1^\prime)B(P_2^\prime).
\numbereq\name{\eqquartic}
$$
The coefficient $<P|{\cal M}|P^\prime>$ is
the inverse di-electron propagator,
given by
$$
<P|{\cal M}|P^\prime>
=(\nu_n^2+\xi_{\vec p}^2)\delta_{PP^\prime}
-{g\over\beta\Omega}\theta(\omega_D-|\xi_{\vec p}|)\theta(\omega_D
-|\xi_{\vec p^\prime}|),
\numbereq\name{\eqdie}
$$
where the second term on the right hand
side represents the leading order 2PI
vertex in accordance with the Hamiltonian (\eqbcs).

The eigenvalues and eigenfunctions of ${\cal M}$ are determined by
$$
(\nu_n^2+\xi_{\vec p}^2)u(P)
-{g\over\beta\Omega}\theta(\omega_D-|\xi_{\vec p}|)
\sum_{\vec p^\prime}\theta(\omega_D
-|\xi_{\vec p^\prime}|)u(P^\prime)=Eu(P).
\numbereq\name{\eqeg}
$$
On writing 
$$
A={g\over\beta\Omega}\sum_{P^\prime}\theta(\omega_D-|\xi_{\vec p}|)u(P)
\numbereq\name{\eqconsta}
$$
we have
$$
u(P)={{A}\over {\nu_n^2+\xi_{\vec p}^2-E}}.
\numbereq\name{\eqwave}
$$
Substituting (\eqwave) back to (\eqconsta), we obtain the secular 
equation for $E$
$$
1={g\over {\beta\Omega}}\sum_P
{{\theta(\omega_D-|\xi_{\vec p}|)}\over
{\nu_n^2+\xi_{\vec p}^2-E}}.
\numbereq\name{\eqsecular}
$$
The pairing mode corresponds to the eigenvalue that
vanishes at $T_C$. For $T$, 
sufficiently close to $T_C$, we may expand
the r. h. s. of (\eqsecular)
to first order in $E$, i.e.
$$
\eqalign{
1&={g\over {\beta\Omega}}
\sum_P{{\theta(\omega_D-|\xi_{\vec p}|)}\over
{\nu_n^2+\xi_{\vec p}^2}}+{{g E}\over {\beta\Omega}}\sum_P
{{\theta(\omega_D-|\xi_{\vec p}|)}\over {(\nu_n^2+\xi_{\vec p}^2)^2}}\cr
&={{gk_F^2}\over {2\pi^2v_F}}
\Big(\ln{{2\omega_D}\over {\pi k_BT}}+\gamma\Big)
+g{{7\zeta(3)k_F^2}\over {16\pi^2v_F(k_BT)^2}}E\cr}
\numbereq\name{\eqxeop}
$$
with $k_F$ the Fermi momentum, $v_F$
the Fermi velocity and $\gamma$ the
Euler constant. At the transition temperature $T=T_c$ the
following relation holds,
$$
1={{gk_F^2}\over {2\pi^2v_F}}
\Big(\ln{{2\omega_D}\over {\pi k_BT_C}}+\gamma\Big)
\numbereq\name{\eqbcst}
$$
and $E=0$ according to (\eqxeop). In the
neighbourhood of $T_C$, we
have
$$
E={{8\pi^2}\over {7\zeta(3)}}k_B^2T_C(T-T_C).
\numbereq\name{\eqarxid}
$$
All other modes are nonpairing with
eigenvalues $E_l=\nu_n^2+\xi_{\vec p}^2+
O(g)$. Expanding the order parameter according 
to the eigenfunctions of
${\cal M}$, i.e.
$$
B(P)=\Psi
u(P)+B^{\prime}(P)
\numbereq\name{\eqmdexp}
$$
with $u(P)$ the eigenfunction of the pairing mode
and ${1\over {\beta\Omega}}\sum_{P}u^{\dagger}(P)
B^{\prime}(P)=0$, we find
$$
\Gamma_2=E\Psi^*{\Psi}+\sum_{P}(\nu_{n}^2+{\xi_{\vec p}}^2)
B^{\prime*}(P)B^{\prime}(P).
\numbereq\name{\eqbgamma}
$$
To calculate the quartic term, we switch off
the interaction and obtain the
quartic term of $J$ in $W(J)$. Then we amputate
it with four di-electron
propagators. The result reads
$$
\Gamma_4={{|\Psi|^4}\over {2\beta\Omega}}
\sum_P(\nu_n^2+\xi_{\vec p})^2
|u(P)|^4+...
\numbereq\name{\eqbgamm}
$$
where the terms in ... contain nonpairing modes and are at most
$\sim|\Psi|^3$ as $T \mapsto T_C$. Neglecting these modes, we find
$$
\Gamma(\Psi)=\Omega\Big[{{8\pi^2k_B^2T_C}\over {7\zeta(3)}}
(T-T_C)|\Psi|^2
+{{4\pi^4(k_BT_C)^2}\over {7\zeta(3)k_F^2}}|\Psi|^4\Big].
\numbereq\name{\eqmnui}
$$

If a small momentum $\vec k$ is introduced, only
the eigenvalue of the pairing
mode is modified by an amount comparable to itself. In order
to determine $E$ in this
case, we observe that (\eqsecular) is replaced by
$$
1={g\over {\beta\Omega}}
\sum_P{{\theta(\omega_D-|\xi_{\vec p}|)}\over
{(i\nu_n-\xi_{\vec p_+})(-i\nu_n-\xi_{\vec p_-})-E}}.
\numbereq\name{\eqseculark}
$$
where $\vec p_\pm={{\vec k}\over {2}}\pm\vec p$. For
the pairing mode, we expand
the r. h. s. to the first order in $E$ and to
the second order in $\vec k$.
After some manipulations, we end up with
$$
E\simeq{{8\pi^2(k_BT_C)^2}\over {7\zeta(3)}}
{{T-T_C}\over {T_C}}+{1\over 6}
v_F^2\vec k^2.
\numbereq\name{\eqinho}
$$
On writing $\Psi_{\vec k}u_{\vec k}(P)$ for
the order parameter at a nozero
$\vec k$ with $u_{\vec k}(P)$ being the pairing mode
at $\vec k\neq 0$, the free
energy functional becomes
$$
\Gamma(\Psi)=\Omega\Big[\Big[{1\over 6}v_F^2\vec k^2
+{{8\pi^2(k_BT_C)^2}\over {7\zeta(3)}}
{{T-T_C}\over {T_C}}\Big]|\Psi_{\vec k}|^2
+{{8\pi^4v_F(k_BT_C)^2}\over {7\zeta(3)k_F^2}}|\Psi_{\vec k}|^4\Big]
\numbereq\name{\eqinhogl}
$$
By summing over $\vec k$ and introducing
$$
\Phi(\vec r)=\sqrt{{2\mu}\over {3}}\int{{d^3\vec k}\over {(2\pi)^3}}
\Psi_{\vec k}e^{i\vec k\cdot\vec r},
\numbereq\name{\eqmean}
$$
we obtain, for an inhomogeneous order parameter, that
$$
\Gamma=\int d^3\vec r\Big[{1\over 2m}\vec\nabla\Phi^*\vec\nabla\Phi
+a|\Phi|^2+{1\over 2}b|\Phi|^4\Big]
\numbereq\name{\eqgfli}
$$
with coefficients
$$
a={{12\pi^2}\over {7\zeta(3)\mu}}k_B^2T_C(T-T_C)
\numbereq\name{\eqcoeffa}
$$
and
$$
b={{36\pi^4v_F}\over {7\zeta(3)}}{{(k_BT_C)^2}\over {\mu^2k_F^2}},
\numbereq\name{\eqcoeffb}
$$
which is in agreement with the result obtained by Gorkov [\gorkov].

\newsection Appendix B.

In order to determine the vertex contribution to
the coefficient $c^{\prime}(T_c,\mu)$ in eq. (\eqeigenA),
we approximate ${\bar\Gamma}_{A, \vec k}(P|P^\prime)$ by considering
only the one magnetic gluon exchange, with both incoming
and outgoing quark momenta near the Fermi surface, i. e.
$$
{\bar\Gamma}_{A, \vec k}(P|P^\prime)\simeq -{i\over {g^2}}
(1+{1\over N})D_{M}({\vec q}, \omega){\bar u}({\vec p}_{+}^{\prime})
{\gamma_i}u({\vec p}_{+}){\bar u}({\vec p}_{-}^{\prime})
{\gamma_j}u({\vec p}_{-})({\delta_{ij}}-{\hat q}_{i}
{\hat q}_{j})
\numbereq\name{\eqacioz}
$$
where $u(\vec p)$
is a solution of the Dirac equation
$(\gamma_4p-i\vec\gamma\cdot\vec p)u(\vec p)=0$
$u^\dagger(\vec p)u(\vec p)=1$, and
$$
{\vec p}_{\pm}={{\vec k}\over 2}\pm {\vec p}, \quad
{\vec p}_{\pm}^{\prime}={{\vec k}\over 2}\pm {\vec p}^{\prime},
\quad {\vec q}={\vec p}-{\vec p}^{\prime}
\numbereq\name{\eqarabia}
$$
with $p \sim p^{\prime} \sim {\mu}$, and $D_{M}({\vec q}, \omega)$
contains the HDL resummation. With the aid of the Dirac equation,
(\eqacioz) can be reduced to
$$
{\bar\Gamma}_{A, \vec k}(P|P^\prime)\simeq -{i\over {g^2}}
(1+{1\over N})D_{M}({\vec q}, \omega)[ V-{{({\vec p}_{+}^{\prime}
-{\vec p}_{+})({\vec p}_{-}^{\prime}
-{\vec p}_{-})}\over {{\vec q}^2}}S]
\numbereq\name{\eqarvitas}
$$
where
$$
\eqalign{
V=&-u^\dagger({\vec p}_{+}^{\prime}){\vec\alpha}
u({\vec p}_{+})u^\dagger({\vec p}_{-}^{\prime}){\vec\alpha}
u({\vec p}_{-})\cr
S=&u^\dagger({\vec p}_{+}^{\prime})
u({\vec p}_{+})u^\dagger({\vec p}_{-}^{\prime})
u({\vec p}_{-})\cr}
\numbereq\name{\eqretro}
$$
with $\vec\alpha=\left(\matrix{0&{\vec\tau}\cr{\vec\tau}&0\cr}\right)$
and $\vec\tau$ the $2\times 2$ Pauli matrices. Since
$u({\vec p})$ is also an eigenstate of $\gamma_5=
\left(\matrix{0&-I\cr-I&0\cr}\right)$ we have
$$
u({\vec p})={1\over {\sqrt 2}}{\pmatrix{{\phi(\vec p)}\cr {\pm}
{\phi(\vec p)}}}
\numbereq\name{\eqpmatrix}
$$
with ${\vec\tau}{\hat p}{\phi}(\vec p)={\phi}(\vec p)$.
Using the identity $({\tau}^j)_{\alpha\beta}
({\tau}^j)_{\rho\lambda}=2{\delta}_{\alpha\lambda}
{\delta}_{\beta\rho}-{\delta}_{\alpha\beta}{\delta}_{\rho\lambda}$
we find that
$$
V=S-2S^{\prime}, \qquad S^{\prime}={\phi}^{\dagger}
({\vec p}_{+}^{\prime}){\phi}
({\vec p}_{-}){\phi}^{\dagger}
({\vec p}_{-}^{\prime}){\phi}({\vec p}_{+}).
\numbereq\name{\eqkhuri}
$$
In the collinear  limit, ${\vec q}\mapsto 0, {\vec p}_{\pm}
\mapsto {\vec p}_{\pm}^{\prime}$, we have $S\mapsto 1$ and
$$
S^{\prime}\mapsto {1\over 2}(1+{\hat p}_{+}{\hat p}_{-})
\simeq {{k^2}\over {p^2}}[1-({\hat p}{\hat k})^2].
\numbereq\name{\eqviagra}
$$
Therefore, a net momentum $\vec k$ shifts the kernel
$K(P|P^{\prime})$ by an amount
$$
{\delta}^{\prime\prime}K(P|P^{\prime})\sim -K(P|P^{\prime})
{{k^2}\over {4p^{\prime2}}}[1-({\hat p}^{\prime}{\hat k})^2].
\numbereq\name{\eqwerop}
$$
Comparing ${\delta}^{\prime\prime}K(P|P^{\prime})$
with the second term of ${\delta}^{\prime}K(P|P^{\prime})$
in (\eqareano), we notice that although the former
differs by a factor of
${\nu_n}^2+(p-\mu)^2$ in the
denominator in comparison with the latter,
it compensates by the factor of $p^2 \sim \mu^2$
which appears in the denominator.
As the sum over Matsubara energies will pick up
the infrared logarithm, $\ln{1\over {\epsilon}}$,
we expect that the contributuion of
${\delta}^{\prime\prime}K(P|P^{\prime})$
to $c^{\prime}(T_c,\mu)$ is suppressed relative to that of
of ${\delta}^{\prime}K(P|P^{\prime})$ by a factor of order
$$
\Big ( {{{k_B}T_c}\over {\mu}} \Big )^2 \ln{{\mu}\over
{{k_B}T_c}} \ll 1
\numbereq\name{\eqfinal}
$$
and therefore can be ignored. 

\newsection Appendix C.

By applying the variational principle to the free
energy functional we obtain the equation of motion
for the special case of $\Phi_l={\vec A}_l=0$ for
$l\ne 0,8$ and ${\vec{\cal V}}=0$,
$$
\eqalign{
&{\vec\nabla}\times({\vec\nabla}\times{\vec V})={\vec J}\cr
&-({\vec\nabla}-2i{\lambda}{\vec V})^2{\phi}+4a{\phi}
+2b(2|{\phi}|^2+|{\chi}|^2){\phi}=0\cr
&-({\vec\nabla}+i{\lambda}{\vec V})^2{\chi}+4a{\chi}
+2b(|{\phi}|^2+{3\over 2}|{\chi}|^2){\chi}=0\cr}
\numbereq\name{\eqmanny}
$$
with the current vector given by
$$
{\vec J}=i{\lambda}\Big [ -2{\phi}^{\star}({\vec\nabla}
-2i{\lambda}{\vec V}){\phi}+{\chi}^{\star}
({\vec\nabla}+i{\lambda}{\vec V}){\chi}\Big ]+ \cdots,
\numbereq\name{\eqjoey}
$$
and the $\cdots$ indicate the complex conjugate. For a vortex
filament parallel to ${\hat\zeta}$, we make the following
ansatz for the solution in cylindrical coordinates
$$
{\phi}=u(\rho)e^{2ni{\cal\theta}}, \qquad
{\chi}=v(\rho)e^{-ni{\cal\theta}}, \qquad {\vec V}
=V(\rho){\hat{\cal\theta}}
\numbereq\name{\eqretoric}
$$
The set of equations (\eqmanny) is reduced to a set
of ordinary non-linear differential equations
$$
\eqalign{
&-{d\over {d\rho}}[{1\over {\rho}}{d\over {d\rho}}({\rho}V)]=
2{\lambda}({n\over {\rho}}-{\lambda}V)(4u^2+v^2) \cr
&-{{d^2u}\over {d{\rho}^2}}-{1\over {\rho}}{{du}\over {d\rho}}
+4({n\over {\rho}}-{\lambda}V)^2u+4au+2b(2u^2+v^2)u=0\cr
&-{{d^2v}\over {d{\rho}^2}}-{1\over {\rho}}{{dv}\over {d\rho}}
+({n\over {\rho}}-{\lambda}V)^2v+4av+2b(u^2+
{3\over 2}v^2)v=0\cr}
\numbereq\name{\eqwerion}
$$
If the color and the flavor were locked everywhere,
we would have $v={\sqrt 2}u$.
The last two equations would then imply that
$$
V={n\over {\lambda\rho}}
\numbereq\name{\eqgauge}
$$
which is a pure gauge for $\rho\ne 0$, as the total flux
has been fixed by the phase of $\phi$ and $\chi$ in (\eqretoric).
A complete color-flavor locking would squeeze all flux in the symmetry
axis of the filament and would make the magnetic field energy
per length infinite.

The next example is the equilibrium between the super
phase and the normal phase under an external $V$-field,
the classical problem of the surface tension of the
interface which was worked out by Ginzburg and Landau
for electronic superconductors \ref{\landau}{V. L. Ginzburg
and L. D. Landau, \ussr20 (1950) 1064.}.
With the interface
parallel to the $y-\zeta$ plane, the solution ansatz reads
$\phi=u(x)$, $\chi=v(x)$ and ${\vec V}=V(x){\hat\zeta}$,
where $u, v$ are real and positive. The set of equations (\eqmanny)
reduces to
$$
\eqalign{
&{{d^2V}\over {dx^2}}=2{\lambda}^2V(4u^2+v^2) \cr
-&{{d^2u}\over {dx^2}}+4{\lambda}^2V^2u
+4au+2b(2u^2+v^2)u=0\cr
-&{{d^2v}\over {dx^2}}+{\lambda}^2V^2v
+4av+2b(u^2+{3\over 2}v^2)v=0\cr}
\numbereq\name{\eqfinal}
$$
An inspection of the last two equations shows that a
complete color-flavor locking, $v={\sqrt 2}u$ can not
implemented unless $V=0$.

\immediate\closeout1
\bigbreak\bigskip

\line{\twelvebf References. \hfil}
\nobreak\medskip\vskip\parskip

\input refs

\vskip 1 in
\input epsf
\centerline{\epsffile{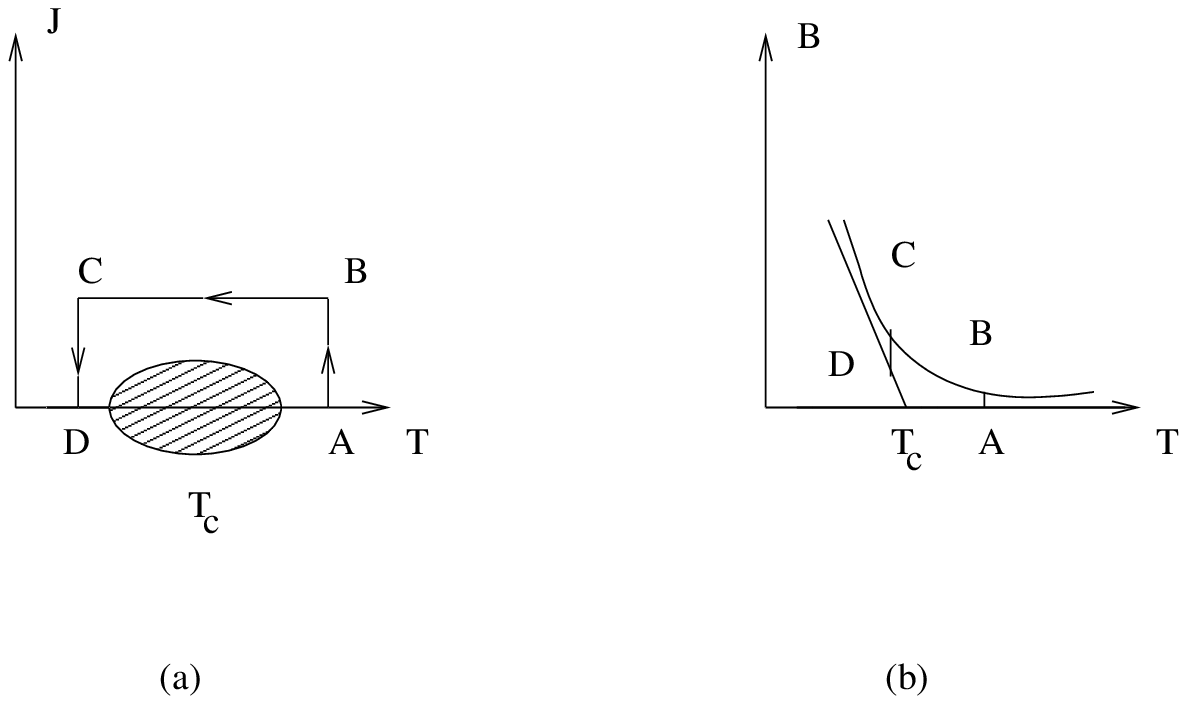}}
\medskip
\noindent
Figure 1
\medskip
The continuation from the normal phase to the super phase
\smallskip
a) The continuation path on $T-J$ plane
\smallskip
b) The order parameter versus temperature

\vskip 1 in
\input epsf
\centerline{\epsffile{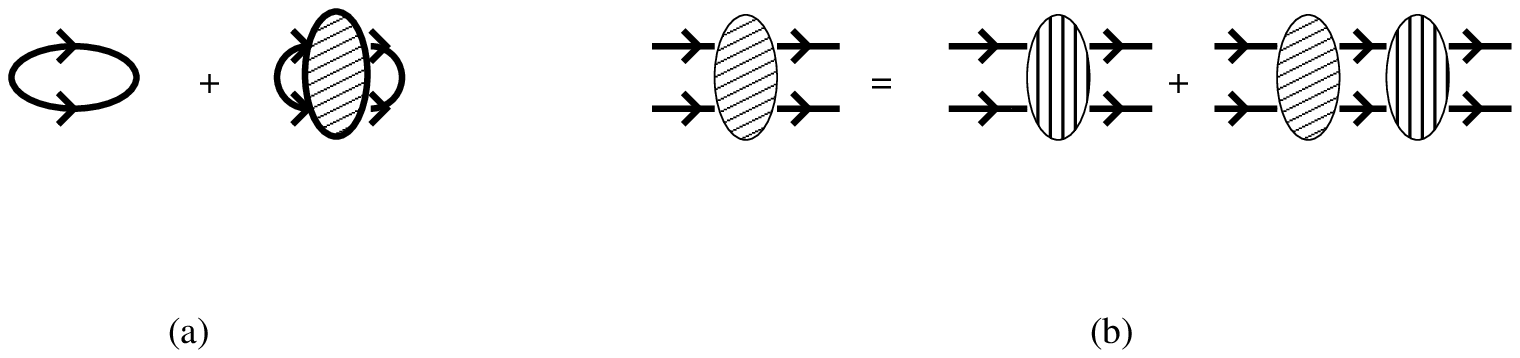}}
\medskip
\noindent
Figure 2
\medskip
a) Diagram of a quark pair propagator. The thick black
line indicates the full quark pair propagator.
\smallskip
b) Dyson-Schwinger equation for the proper vertex of
diquark scattering, where the vertically shaded vertex stands for
two quark irreducible diagrams.

\vskip 1 in
\input epsf
\centerline{\epsffile{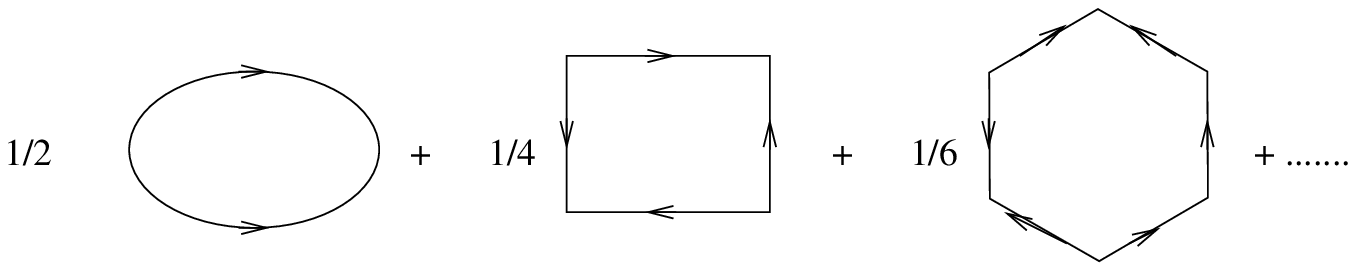}}
\medskip
\noindent
Figure 3
\medskip
The diagrammatic expansion of W(J) at g=0.

\vfil\end

\bye